\documentclass[pra,aps,twocolumn,groupedaddress]{revtex4}
\usepackage{amssymb,amsmath,amsfonts,epsfig,graphicx,latexsym,bm,color}

\usepackage{subfigure}

\newcommand{\beq}{\begin{eqnarray}}
\newcommand{\eeq}{\end{eqnarray}}
\newcommand{\bem}{\begin{pmatrix}}
\newcommand{\eem}{\end{pmatrix}}
\newcommand{\nn}{\nonumber}
\newcommand{\hb}{\hbar}

\newcommand{\f}{\frac}
\newcommand{\rfs}[1]{Eq.~(\ref{#1})}

\newcommand{\tb}[1]{\textbf{#1}}
\newcommand{\tr}[1]{\textrm{#1}}
\newcommand{\ro}[1]{\sqrt{#1}}
\newcommand{\ex}[1]{\langle #1 \rangle}
\def\nn{\nonumber}

\def\e{\epsilon}
\def\d{\delta}
\def\D{\Delta}

\def\Om{\Omega}
\def\p{\phi}
\def\B{\beta}

\def\A{\alpha}

\def\s{\sigma}

\def\lt{\left}
\def\rt{\right}

\def\da{\downarrow}
\def\ua{\uparrow}

\begin{document}
\title{Competing Superconducting States for Ultracold Atoms in Optical Lattices with Artificial Staggered Magnetic Field}
\author{Lih-King Lim$^{1}$, Achilleas Lazarides$^{1}$, Andreas Hemmerich$^{2}$, and C. Morais Smith$^{1}$}
\affiliation{$^{1}$Institute for Theoretical Physics, Utrecht University,
Leuvenlaan 4, 3584 CE Utrecht, The Netherlands}
\affiliation{$^{2}$Institut f\"{u}r Laser-Physik, Universit\"{a}t Hamburg,
Luruper Chaussee 149, 22761 Hamburg, Germany}
\date{\today}

\begin{abstract}
We study superconductivity in an ultracold Bose-Fermi mixture loaded into a square optical lattice subjected to a staggered flux. While the bosons form a superfluid at very low temperature and weak interaction, the interacting fermions experience an additional long-ranged attractive interaction mediated by phonons in the bosonic superfluid. This leads us to consider a generalized Hubbard model with on-site and nearest-neighbor attractive interactions, which give rise to two competing superconducting channels. We use the Bardeen-Cooper-Schrieffer theory to determine the regimes where distinct superconducting ground states are stabilized, and find that the non-local pairing channel favors a superconducting ground state which breaks both the gauge and the lattice symmetries, thus realizing unconventional superconductivity. Furthermore, the particular structure of the single-particle spectrum leads to unexpected consequences, for example, a dome-shaped superconducting region in the temperature versus filing fraction phase diagram, with a normal phase that comprises much richer physics than a Fermi-liquid. Notably, the relevant temperature regime and coupling strength is readily accessible in state of the art experiments with ultracold trapped atoms. 
\end{abstract}

\maketitle
\date{\today}

\section{Introduction}
Superconductivity in low dimensional systems is more than ever an active field of research. Despite decades of research, one of its holy grail remains to be a thorough understanding of high-temperature superconductivity (high-$T_c$) in the cuprates \cite{Bednorz:86,Bonn:06}. Even though the Bardeen-Cooper-Schrieffer (BCS) theory has been proven to be very successful in addressing ordinary superconducting phenomena, its straightforward application to understand the strongly correlated regime (e.g. in high-$T_c$ superconductors) remains a challenge. 

In this article we argue that, in combination with unconventional single-particle spectra, BCS theory gives rise to phenomena reminiscent of the physics known to occur in strongly correlated systems. We therefore study a ultracold atom system with a Dirac-like spectrum with linear rather than the ordinary quadratic dispersion, arising via a time-reversal symmetry braking term in the Hamiltonian. At first sight, the relevance of Dirac fermions appears to be limited to a relativistic context. However, such fermions do emerge in condensed matter systems under certain circumstances. Examples include high-$T_c$ cuprate superconductors, which exhibit a $d_{x^2-y^2}$ symmetry in the superconducting order parameter, such that fermionic excitations along the nodal lines on the Fermi surface are Dirac-like \cite{Tsuei:00}, and graphene, where the hexagonal crystal lattice gives rise to Dirac-like excitations \cite{Neto:09, Geim:09}. 

Indeed, the recent breakthrough in fabricating sheets of graphene has provided us with a solid-state model of two-dimensional Dirac fermions \cite{Novoselov:04}. The observation of the half-integer quantum Hall effect \cite{Novoselov:05} and Klein tunneling \cite{Young:09, Stander:09} in graphene, for example, are hallmarks of two-dimensional relativistic physics taking place in a condensed matter system. Interesting theoretical works have explored the importance of interaction effects in graphene, where phenomena such as room-temperature superfluidity \cite{Min:08}, an anomalously low shear viscosity in the vicinity of quantum criticality \cite{Muller:09} and novel superconductivity \cite{Neto:07, Black:07} have been predicted. On the experimental front, the anticipated fractional quantum Hall effect has only been observed very recently in an exfoliated graphene sample \cite{Du:09,Bolotin:09}. Nevertheless, it still requires ingenuity to prepare a clean, highly-controllable graphene-based system, in order to fully explore its rich physics.

In ultracold atomic systems, on the other hand, the remarkable progress made in the last decade has allowed for experimental demonstrations of prototypical many-body phenomena, as the superfluid-Mott insulator transition in the Bose-Hubbard model \cite{Fisher:89, Jaksch:98, Greiner:02, Spielman:08} and the crossover between the Bardeen-Cooper-Schrieffer (BCS)-Bose-Einstein condensation (BEC) regimes \cite{Tiesinga:93, Inouye:98, Bartenstein:04,Bourdel:04,Regal:04,Zwierlein:04}. One of the major tasks in the field is the use of optical lattices for systematic emulations of the Hubbard model, a model which is believed to capture the physics of high-$T_c$ superconductors. A particular interesting option in optical lattices is the study of degenerate Bose-Fermi mixtures. The relative ease in tuning the parameters such as the interspecies interaction strength, and the density and mass ratios, offers a wide window for studies confronting theory with precision experiments. Many interesting phenomena have been investigated, which include the prediction of exotic quantum matters, such as supersolids \cite{Buchler:03, Titvinidze:08, Orth:09}, composite fermions \cite{Lewenstein:04}, charge density waves and polaron-like quasi-particles \cite{Mathey:04}. Also more common effects arise, such as the enhancement of different types of fermionic superfluidity mediated by the phonon background provided by the bosons \cite{Illuminati:04,Wang:05}, which would otherwise be difficult to control in other systems. On the experimental side, the ability to control the experimental parameters in Bose-Fermi mixtures in optical lattices is steadily progressing \cite{Gunter:06, Ospelkaus:06, Best:09}.

\begin{figure}[t]
\includegraphics[scale=.54, angle=0, origin=c]{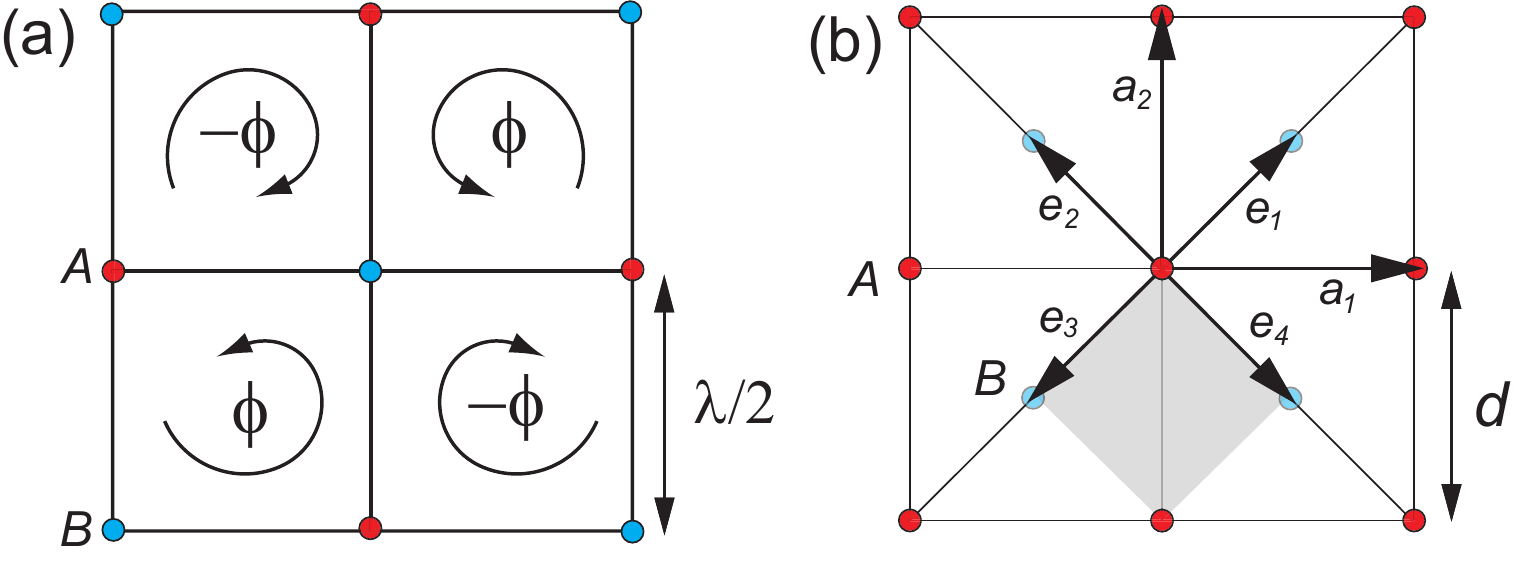}
\caption{\label{Fig.1}(color online) (a) Schematic of the two-dimensional square optical lattice with a spacing of half the laser wavelength $\lambda/2$. Under the presence of a flux $\p$ passing through each plaquette that alternates in sign across the neighboring plaquette -- staggered flux -- it is then convenient to split the lattice into two inequivalent sublattices $\mathcal{A}$ and $\mathcal{B}$.  (b) The Bravais lattice is made of the $\mathcal{A}$-sublattice with a lattice constant $d=\sqrt{2}\lambda/2$. The unit cell consists of a basis, which is defined on the $\mathcal{B}$-sublattice. The four vectors $\tb{e}_l$, with $l=1-4$, connect each $\mathcal{A}$-site to its four nearest-neighboring $\mathcal{B}$-sites.}
\end{figure}

In this work, we study the system of Bose-Fermi mixtures in a two-dimensional optical square lattice that provides an inherent staggered flux. In a recent letter we have shown the emergence of Dirac fermions and the stabilization of various superconducting states for this system \cite{Lim:09}. Here, we present the details of the calculations and extend the work to include a discussion on the competing spin-liquid instability. In addition, we study the re-entrant behavior due to the specific pairing states and the evolution of the Fermi surface as a function of the staggered flux. 

The paper is organized as follows: in Sec. II, we use the Bogoliubov theory to study the mediation effects of the condensed bosons on the fermions in the presence of the staggered flux, thus providing a method to realize an extended Hubbard interaction. In Sec. III, we develop a mean-field theory to study superconductivity with the model. We first discuss possible pairing states in the separate interaction channels and their associated symmetries. We then construct the free energy and obtain the gap equations. We find a re-entrant behavior, which is unique to the local pairing channel. As for the non-local pairing channel, we compare the transition temperatures of the various possible pairing states. We then discuss the evolution of the Fermi surface as the staggered flux is tuned away from the case of a $\pi$-flux. In Sec. IV, experimental signatures are discussed for the detection of the various superconducting order parameters and the evolution of the Fermi surfaces. Finally, we close the paper with a discussion and conclusions in Sec. V.

\section{Bose-Fermi Mixtures}
We start from the microscopic single-band Bose-Fermi Hubbard Hamiltonian subjected to a staggered flux $\p$ \cite{Lim:08a,Lim:10},
\beq\label{BFmix}
H&=&H_{0,B}+H_{0,F}+H_{BB}+H_{FF}+H_{BF},
\eeq
where
\beq
H_{0,B}&\equiv&-J_B\sum_{\substack{\tb{r} \in \mathcal{A}\\l=1-4}} \biggl( e^{i\p(-1)^l/4}a_{\tb{r}}^{\dag} b_{\tb{r}+\tb{e}_l}+\tr{H.c.}\biggr)\nn\\
H_{0,F}&\equiv&-J \sum_{\substack{\tb{r} \in \mathcal{A}\\l=1-4}}\sum_{\s} \biggl( e^{i\p(-1)^l/4}a_{\tb{r},\s}^{\dag} b_{\tb{r}+\tb{e}_l,\s}+\tr{H.c.}\biggr)\nn\\
H_{BB}&\equiv&\f{U_{BB}}{2}\sum_{\tb{r}\in \mathcal{A}\oplus \mathcal{B}}n_{\tb{r}}^B\,\,(n_{\tb{r}}^B-1),\nn\\
H_{FF}&\equiv&\f{U_{FF}}{2}\sum_{\tb{r}\in \mathcal{A}\oplus \mathcal{B}}\sum_\s n_{\tb{r},\s}\,\,n_{\tb{r},-\s},\nn\\
H_{BF}&\equiv& U_{BF}\sum_{\tb{r}\in \mathcal{A}\oplus \mathcal{B}}\sum_\s n_{\tb{r}}^B\,\,n_{\tb{r},\s}.
\eeq
Here, $J_{B}$ and $J$ are the hopping amplitudes for bosons and fermions, respectively, and $U_{BB}$, $U_{FF}$ and $U_{BF}$ are the boson-boson, fermion-fermion and boson-fermion on-site interactions, respectively. The presence of the staggered magnetic field leads to the appearance of two inequivalent $\mathcal{A}$- and $\mathcal{B}$-sublattices, see Fig. \ref{Fig.1}(a). The operators $a_{\tb{r}}$ and $b_{\tb{r}+e_l}$ are the bosonic annihilation operators acting on site $\tb{r}$ and $\tb{r}+\tb{e}_l$ of the $\mathcal{A}$- and $\mathcal{B}$-sublattices, respectively, and $a_{\tb{r},\s}$ and $b_{\tb{r}+e_l,\s}$ are the corresponding fermionic annihilation operators with spin $\s$. Finally, $n_{\tb{r}}^B$ and $n_{\tb{r},\s}$ are the boson and fermion number operators on site $\tb{r}$, respectively.

We write the grand canonical partition function $Z$ in the functional-integral representation
\beq\label{action}
Z=\int \mathcal{D}a\mathcal{D}a^*\mathcal{D}a_\s\mathcal{D}a_\s^* \exp\biggl\{-\f{1}{\hb}(S_B+S_F+S_I)\biggr\}
\eeq
where the bosonic action $S_B$ is given by
\beq
S_B=\int_0^{\hb \B} d\tau \biggl[ \sum_{i\in \mathcal{A}\oplus\mathcal{B}} a_i^*(\tau)(\hb\partial_\tau-\mu)a_i(\tau)+H_{0,B}+H_{BB}\biggl],\nn
\eeq
the fermionic action $S_F$ is given by
\beq
S_F&=&\int_0^{\hb \B} d\tau \biggl[ \sum_{i\in \mathcal{A}\oplus\mathcal{B}}\sum_\s a_{i,\s}^*(\tau)(\hb\partial_\tau-\mu_\s)a_{i,\s}(\tau)+H_{0,F}\nn\\&&+H_{FF}\biggl],\nn
\eeq
and the boson-fermion interaction term $S_{I}$ is given by
\beq
S_I=H_{BF}.\nn
\eeq
Here, $\mu$ is the chemical potential for bosons, and $\mu_\s$ the one for fermions with spin $\s$. The inverse thermal energy at temperature $T$ is given by $\B=1/k_B T$.

We focus on the weakly interacting regime of the atomic bosons where they form a Bose-Einstein condensate (BEC) in the lattice. Such regime can be described accurately within the Bogoliubov theory.
% with condensate fraction $\tilde{n}$ and healing length $\xi\equiv (J_B/\tilde{n} U_{BB})^{1/2}$.
Before we apply the theory, we need to identify the condensation mode for the bosons, since the presence of the staggered flux can give rise to distinct BECs.

We first note that the single-particle term $H_{0,B}$ can be diagonalized in momentum space with the canonical transformation
\beq
a_{\tb{k}}(\tau)&=&\f{1}{\sqrt{2}}\f{\e_{\tb{k}}^*}{|\e_{\tb{k}}|}(-\A_{\tb{k}}(\tau)+\B_{\tb{k}}(\tau)),\nn\\
b_{\tb{k}}(\tau)&=&\f{1}{\sqrt{2}}(\A_{\tb{k}}(\tau)+\B_{\tb{k}}(\tau)),
\eeq
where 
\beq\label{con}
\e_{\tb{k}} &=&4J\biggr[\cos\biggl(\f{\phi}{4}\biggr)\cos\biggl(\f{k_x d}{2}\biggr)\cos\biggl(\f{k_y d}{2}\biggr)\nn\\&&-i\sin\biggl(\f{\phi}{4}\biggr) \sin\biggl(\f{k_x d}{2}\biggr) \sin\biggl(\f{k_y d}{2}\biggr)\biggr]
\eeq
is the lattice dispersion, and the new operators $\A_{\tb{k}}(\tau),\B_{\tb{k}}(\tau)$ correspond to the upper and lower energy band states. The bosonic operator defined on the sublattice $\mathcal{A}$ can then be written as
\beq
a_i(\tau)&=&\f{1}{\sqrt{N_A}} \sum_{\tb{k}\in \tr{1BZ}}a_\tb{k}(\tau) e^{i \textbf{k}\cdot \tb{r}_i}\nn\\
&=&\f{1}{\sqrt{ N}}\sum_{\tb{k}\in \tr{1BZ}}\lt[\f{\e_\tb{k}^*}{|\e_{\tb{k}}|}\B_\tb{k}(\tau)-\f{\e_\tb{k}^*}{|\e_{\tb{k}}|}\A_\tb{k}(\tau) \rt] e^{i \tb{k}\cdot \tb{r}_i},\nn
\eeq
where the lattice momentum $\tb{k}$ is defined in the first Brillouin zone $(1BZ)$ and the canonical transformation is used. Note that the total number of lattice sites is $N=2N_A$. To simplify the following calculations, we map the upper band operator $\A_\tb{k}(\tau)$ in the first Brillouin zone to the second Brillouin zone $(2BZ)$ which is then denoted by $\B_\tb{k}(\tau)$. Since the transformation coefficient changes an overall sign in the second Brillouin zone, we have
\beq
a_i(\tau)&=&\f{1}{\sqrt{ N}}\sum_{\tb{k}\in \tr{1BZ} \oplus\tr{2BZ}}\f{\e_\tb{k}^*}{|\e_{\tb{k}}|} \B_{\tb{k}}(\tau) e^{i \tb{k}\cdot \tb{r}_i}\nn\\&\equiv& \f{1}{\sqrt{ N}}\sum_{\tb{k}\in \tr{1BZ} \oplus\tr{2BZ}}g_{\tb{k}} \B_{\tb{k}}(\tau) e^{i \tb{k}\cdot \tb{r}_i}.
\eeq
Similarly, for the bosonic operators on the $\mathcal{B}$ sublattice, we have
\beq
b_i(\tau)=\f{1}{\sqrt{N}}\sum_{\tb{k}\in \tr{1BZ} \oplus\tr{2BZ}} \B_\tb{k}(\tau) e^{i \tb{k}\cdot (\tb{r}_i+\tb{e}_1 )}.
\eeq
We may now identify the condensation mode $\tb{k}_0$ as the single-particle state with the lowest energy, and perform the $c$-number substitution for the condensate field $\B_{\tb{k}_0}(\tau)\rightarrow\sqrt{N_0}$ as follows
\beq
a_i(\tau)&=&g_{\tb{k}_0}\sqrt{n_0}e^{i \tb{k}_0\cdot \tb{r}_i}+\f{1}{\sqrt{N}}\sum_{\tb{k}}' g_{\tb{k}} \B_{\tb{k}}(\tau) e^{i \tb{k}\cdot \tb{r}_i},
\nn\\
b_i(\tau)&=&\sqrt{n_0}e^{i \tb{k}_0\cdot (\tb{r}_i+\tb{e}_1)}+\f{1}{ \sqrt{ N}}\sum_{\tb{k}}' \B_\tb{k}(\tau) e^{i \tb{k}\cdot (\tb{r}_i+\tb{e}_1)},\nn
\eeq 
$n_0=N_0/N$ defines the condensate density. The prime in the momentum summation means that the $\tb{k}=\tb{k}_0$ term is omitted. 

By expanding also the fields in Matsubara frequencies
\beq
\B_\tb{k}(\tau)=\f{1}{\sqrt{\hb \B}}\sum_{m} e^{-i\omega_m\tau}\B_{\tb{k}}(\omega_m),
\eeq
with $\omega_m=2\pi m/\hb \B$, we make the Bogoliubov approximation in the action $S_B+S_I$ where the bosonic fluctuation field $\B_\tb{k}(\tau)$ is kept to the quadratic order to obtain (see Appendix)
\begin{widetext}
\beq\label{fluc}
S_B+S_I&=&-\f{1}{2}N \hb \B U_{BB} n_0^2\nn\\&&+\f{1}{2}\sum_{\tb{k},m}
\bem
\B_{\tb{k}}(\omega_m)\\\B_{-\tb{k}}^\dag(-\omega_m)
\eem^\dag
\bem
 -i\hb \omega_m+E_{\tb{k}}-E_{\tb{k}_0}+U_{BB} n_0& \f{1}{2}M(\tb{k},-\tb{k},\tb{k}_0,\tb{k}_0)U_{BB} n_0\\
 \f{1}{2}M(\tb{k}_0,\tb{k}_0,\tb{k},-\tb{k}) U_{BB} n_0& i\hb \omega_m+E_\tb{k}-E_{\tb{k}_0}+U_{BB} n_0
\eem
\bem
\B_{\tb{k}}(\omega_m)\\\B_{-\tb{k}}^\dag(-\omega_m)
\eem\nn\\
&&+U_{BF}\sqrt{\f{n_0}{ N}}\sum_{\tb{k},m} \bem
J_{\tb{k}}(\omega_m)\\J_{-\tb{k}}^\dag (-\omega_m)
\eem^\dag \cdot\bem
\B_{\tb{k}}(\omega_m)\\\B_{-\tb{k}}^\dag(-\omega_m)
\eem\nn\\
&&\equiv-\f{1}{2}N \hb \B U_{BB} n_0^2+\f{1}{2}\sum_{\substack{\tb{k}\in \tr{1BZ} \oplus\tr{2BZ},\\m}}\lt[ \vec{\phi}^{\,\,\dag}\cdot \lt(-\hb \tb{G}^{-1}\rt)\cdot\vec{\phi}+\vec{J}^{\,\,\dag}\cdot \vec{\phi}+\vec{\phi}^{\,\,\dag}\cdot \vec{J} \rt]
\eeq
where
\beq
\vec{\phi}\equiv
\bem
\B_{\tb{k}}(\omega_m)\\\B_{-\tb{k}}^\dag(-\omega_m)
\eem,\tr{\ \ \ \ }
\vec{J}\equiv
U_{BF}\sqrt{\f{n_0}{N}}
\bem
J_{\tb{k}}(\omega_m)\\J_{-\tb{k}}^\dag (-\omega_m)
\eem,
\eeq
with the source term containing the fermionic fields given by
\beq
J_{\tb{k}}(\omega_m)&\equiv& g_{\tb{k}_0}g^*_{\tb{k}} \sum_{i\in\mathcal{A}}\sum_\s n_{i,\s}  e^{i(\tb{k}_0-\tb{k})\cdot\tb{r}_i}+\sum_{i\in\mathcal{A}}\sum_{\s}n_{i,\s} e^{i(\tb{k}_0-\tb{k}) \cdot(\tb{r}_i+\tb{e}_1)},
\eeq
\end{widetext}
and $E_\tb{k}=\mp |\e_\tb{k}|$ for $\tb{k}\in \tr{1BZ}  (\tr{2BZ})$, and $M(\tb{k}_1,\tb{k}_2,\tb{k}_3,\tb{k}_4)\equiv 1+g^*_{\tb{k}_1}g^*_{\tb{k}_2}g_{\tb{k}_3}g_{\tb{k}_4}$.
Several symmetry properties have been employed: $g_\tb{k}=g_{-\tb{k}}$ and $\tb{k}_0\equiv\{\pm \tb{k}_0\}$, in the sense that for $\tb{r}=\A\tb{d}_1+\gamma\tb{d}_2$ one gets $\exp[i\tb{k}_0\cdot \tb{r}]\equiv \exp[i\pi(\pm\A\pm\gamma)]$. We have chosen the bosonic chemical potential to obey the Hugenholtz-Pines form, which depends also on the fermion mean-field density $\tilde{n}_\s$,
\beq
\mu=E_{0}+U_{BB} n_0+U_{BF}\sum_\s \tilde{n}_\s,
\eeq
so that the bosonic fluctuation field remains massless (the Goldstone mode). The action \rfs{fluc} is now at most of quadratic order in the bosonic fluctuation field $\vec{\phi}$ and we can thus integrate it out analytically \cite{Stoof:00,Viverit:00}.
%%\beq
%%\int {\cal D}\vec{\phi}\, \exp \biggl\{ -\f{1}{\hb}( S_B+S_I )\biggr\}
%%=\exp\biggl(-\f{1}{2\hb^2} \sum_{\tb{k}\in \tr{1BZ} \oplus\tr{2BZ}}
%%\vec{J}^{\,\,\dag} \cdot \tb{G} \cdot \vec{J}   \biggr).
%%\eeq

Starting from the original action (\ref{action}), by performing the Bogoliubov approximation and integrating out the bosons, we arrive at the effective action for the fermions, 
\beq
S&=&S_F+\f{1}{2\hb}\sum_{\tb{k}\in \tr{1BZ} \oplus\tr{2BZ}}\sum_m \vec{J}^{\,\,\dag}_{\tb{k}} \cdot \tb{G}_{\tb{k}} \cdot \vec{J}_{\tb{k}}\nn\\&\equiv& S_F+S_{ind}
\eeq
where the induced interaction is given by
\begin{widetext}
\beq
S_{ind}=-\f{U_{BF}^2 n_0}{ 2N}\sum_{\textbf{k},m}\f{Y_{\textbf{k},\textbf{k}_0}(J_\textbf{k}^\dag J_\textbf{k}+J_\textbf{k} J_\textbf{k}^\dag)-\f{1}{2}U_{BB}n_0(\tilde{M} J_\textbf{k}^\dag J_{-\textbf{k}}^\dag+\tilde{M}^*J_{-\textbf{k}} J_\textbf{k})}{(\hb \omega_m)^2+W_{\textbf{k}}}.
\eeq
Here, short hand notations are used for the form factor $\tilde{M}\equiv M(\tb{k},-\tb{k},\tb{k}_0,\tb{k}_0)$, $Y_{\textbf{k},\textbf{k}_0}\equiv (E_\textbf{k}-E_{\textbf{k}_0}+U_{BB}n_0) $ and $W_{\textbf{k}}\equiv Y_{\textbf{k},\textbf{k}_0}^2-|\tilde{M}|^2U_{BB}^2n_0^2/4$. 

Next, we consider the frequency independent component of the induced interaction, a well-studied regime for Bose-Fermi mixtures (e.g., the widely used rubidium-potassium system \cite{RbK}) in the \textit{lattice}. Due to differences in the laser detuning as experienced by the different atomic species, the hopping amplitude can be realized for $J_B\gg J$ so that the consideration of the static limit is justified \cite{Illuminati:04}. By setting $\hb \omega_m=0$ and converting the momentum summation into a momentum integral $(1/N)\sum_\textbf{k}\rightarrow d^2 \int_{1BZ\oplus2BZ} d^2 \textbf{k}/(2\pi)^2$ we get
\beq \label{inducedpot}
S_{int}&=&\f{1}{2}\sum_{\tb{r},\tb{r}'\in \mathcal{A}}\sum_{\s,\s'}V_{\mathcal{A}\mathcal{A}}(\tb{r}-\tb{r}')
\bigl[ n_{\tb{r},\s} n_{\tb{r}',\s'}+n_{\tb{r}+\tb{e}_1,\s} n_{\tb{r}'+\tb{e}_1,\s'}\bigr]+\sum_{\tb{r},\tb{r}'\in\mathcal{A}}\sum_{\s,\s'} V_{\mathcal{A}\mathcal{B}}(\tb{r}-\tb{r}'-\tb{e}_1)n_{\tb{r},\s} n_{\tb{r}'+\tb{e}_1,\s'},
\eeq
where the induced potentials are given by
\beq
V_{\mathcal{A}\mathcal{A}}(\tb{r}-\tb{r}')&=&-U_{BF}^2 n_0 d^2\int\f{d^2\textbf{k}}{(2\pi)^2} \f{2}{W_\textbf{k}}\biggl\{  Y_{\textbf{k},\textbf{k}_0} \cos[\textbf{k}_0\cdot(\tb{r}-\tb{r}')]\cos[\textbf{k}\cdot(\tb{r}-\tb{r}')]\nn\\&&
-U_{BB}n_0\cos[\textbf{k}\cdot(\tb{r}-\tb{r}')]\cos[\tb{k}_0\cdot(\tb{r}+\tb{r}')]\cos^2[\theta_0-\theta_{\tb{k}}] \biggr\}\equiv-\f{4U_{BF}^2}{U_{BB}}\,V_2\biggl( \textbf{k}_4,\f{J_B}{U_{BB}n_0},\f{|\tb{r}-\tb{r}'|}{d}\biggr)
\eeq
and
\beq
V_{\mathcal{A}\mathcal{B}}(\tb{r}-\tb{r}'-\tb{e}_1)&=&-U_{BF}^2 n_0 d^2\int \f{d^2\tb{k}}{(2\pi)^2} \f{4}{W_\tb{k}} \biggl\{ Y_{\textbf{k},\textbf{k}_0}\cos[\textbf{k}\cdot (\tb{r}-\tb{r}'-\tb{e}_1)]\cos[\textbf{k}_0\cdot(\tb{r}-\tb{r}'-\tb{e}_1)-\theta_\textbf{k}+\theta_0]\nn\\
&&
-U_{BB}n_0 \cos[\textbf{k}_0\cdot(\tb{r}+\tb{r}'+\tb{e}_1)]\cos[\theta_0-\theta_\textbf{k}]\cos[\textbf{k}\cdot(\tb{r}-\tb{r}'-\tb{e}_1)]\biggr\}\nn\\&&\equiv-\f{4U_{BF}^2}{U_{BB}}\,V_1\biggl( \textbf{k}_4,\f{J_B}{U_{BB}n_0},\f{|\tb{r}-\tb{r}'-\tb{e}_1|}{d}\biggr).
\eeq\end{widetext}
These are the nonlocal phonon-mediated fermion density-density interaction terms, which take the form of a Yukawa potential  (screened Coulomb) in two dimensions. Mediated by the phonons, a non-local attractive interaction is generated between fermions of all spin states, which fall off on the scale of the healing length $\xi$. Experiments in a bosonic 2D lattice of rubidium atoms by Spielman \textit{et al.} \cite{Spielman:08} show that typical values of $\xi$ are on the order of $d/\sqrt{2}$. For $J_B/U_{BB}n_0=1$, the induced potentials are $V_{\mathcal{A}\mathcal{B}}(\tb{e}_1)\simeq-(4U_{BF}^2/U_{BB})\,0.17$ while $V_{\mathcal{A}\mathcal{A}}(\tb{d}_1)$ is one order of magnitude smaller.

Thus the effective interactions are the renormalized on-site interaction $g_1\equiv -U_{FF}-V_{\mathcal{A}\mathcal{A}}(0)$, and the nearest neighbor interaction $g_2\equiv -V_{\mathcal{A}\mathcal{B}}(\tb{e}_1)\simeq (4U_{BF}^2/U_{BB})\,0.17$. Note that the boson-boson interaction $U_{BB}$ is taken to be positive for the stability of the Bose gas and thus, the induced interaction is always attractive. We write the effective interaction for the fermionic Hamiltonian as
\beq
H_{int}&=&-\f{g_1}{2}\sum_{\tb{r}\in \mathcal{A}\oplus \mathcal{B}}\sum_{\s}n_{\tb{r},\s}n_{\tb{r},-\s}\nn\\&&-g_2\sum_{<\tb{r},\tb{r}'>}\sum_{\s,\s'} n_{\tb{r},\s}n_{\tb{r}',\s'}\nn\\
&\equiv& H_1+H_2
\eeq
Notice that since $U_{FF}$ can be tuned by changing the scattering length $a_s$ via Feshbach resonances, the relative strength $g_1/g_2$ can be tuned in a straightforward manner. Furthermore, the dependence of the induced potentials on the parameter $(J_B/U_{BB} n_0)$ can also be taken as an extra tuning parameter; both techniques thus allow for an independent control of both $g_1$ and $g_2$.

\section{Novel Superconductivity}\label{sec2}
In the last section, we studied the attractive mediation effect of the bosonic superfluid on the fermions, which is shown to extend to the nearest-neighbor sites. The effective Hamiltonian describing the interacting fermions 
\beq\label{interactingHAmiltonian}
H=H_{0,F}+H_1+H_2
\eeq
%\beq\label{interactingHAmiltonian}
%H&=&-J \sum_{\substack{\tb{r} \in \mathcal{A}\\l=1-4}}\sum_{\s} \biggl( e^{i\p(-1)^l/4}a_{\tb{r},\s}^{\dag} b_{\tb{r}+\tb{e}_l,\s}+\tr{H.c.}\biggr)\nn\\&&-\f{g_1}{2}\sum_{\tb{r}\in \mathcal{A}\oplus \mathcal{B}}\sum_{\s}n_{\tb{r},\s}n_{\tb{r},-\s}-g_2\sum_{<\tb{r},\tb{r}'>}\sum_{\s,\s'} n_{\tb{r},\s}n_{\tb{r}',\s'}\nn\\
%&\equiv& H_0+H_1+H_2
%\eeq
provides the basis for the study of superconducting instabilities due to the competing on-site and the nearest-neighbor attractive interactions.
%By using BCS mean-field theory, we consider two types of superconducting instabilities and later, we justify the use of the mean-field approach by showing that the required coupling strength to induce superconductivity is within the weak-coupling regime.
%Because the order parameters $\D_i$ introduced for the various channels enter together with the corresponding coupling strength in the mean-field Hamiltonian, the absence of order $\D_i=0$ also means the absence of dependence on the coupling strength in the free energy. This is an artifact of the mean-field approach, where the Hartree correction is neglected. However, to look for superconductivity at the mean-field level, this is generally the case and the full free energy function has to be verified.
For clarity of presentation, we first identify the various order parameters that are favored by the interaction terms and carry out the analysis for the different superconducting channels independently. Then, we construct the full phase diagram in the presence of both couplings.

\subsection{Pairing Hamiltonian}
For the on-site attractive interaction $g_1>0$, we consider an on-site spin-singlet pairing with the following superconducting order parameter  
\beq
\D_{1}&\equiv&-\f{2g_1}{N}\sum_{i\in\mathcal{A}}\ex{a_{i,\da}a_{i,\ua}}=-\f{2g_1}{N}\sum_{i\in\mathcal{B}}\ex{b_{i,\da}b_{i,\ua}}.\nn
\eeq
By performing a mean-field decoupling in $H_1$ with respect to this superconducting order parameter and keeping fluctuations up to first order, we arrive at the following mean-field Hamiltonian 
\beq
H_{1,MF}&\simeq&\f{N|\D_1|^2}{g_1}+\sum_{\textbf{k}}\biggl[ \D_{1}^\dag\, a_{\textbf{k},\da} a_{-\textbf{k},\ua} +\D_{1}^\dag \,b_{\textbf{k},\da} b_{-\textbf{k},\ua}\nn\\&&+\tr{H.c.}\biggr].
\eeq

For the nearest-neighbor attractive interaction $g_2>0$, we consider the following decomposition:
\beq\label{SC2}
H_2&=&-g_2\biggl\{\sum_{<i,j>}\biggl[(a_{i,\ua}^\dag b_{j,\da}^\dag-a_{i,\da}^\dag b_{j,\ua}^\dag)(a_{i,\da}b_{j,\ua}-a_{i,\ua} b_{j,\da})\nn\\&&+\f{1}{2}(a_{i,\ua}b_{j,\ua}^\dag+a_{i,\da}b_{j,\da}^\dag)(a_{i,\ua}^\dag b_{j,\ua}+a_{i,\da}^\dag  b_{j,\da})\nn\\&&
+\f{1}{2}(a_{i,\ua}^\dag b_{j,\ua}+a_{i,\da}^\dag b_{j,\da})(a_{i,\ua} b_{j,\ua}^\dag+a_{i,\da}  b_{j,\da}^\dag)\biggr]\nn\\&&+\f{1}{2}\sum_{i,\s}\biggl(a_{i,\s}^\dag a_{i,\s}+b_{i,\s}^\dag b_{i,\s}\biggr)\biggr\}.
\eeq
The first term may give rise to a non-local pairing correlation involving two nearest-neighbor sites with a spin-singlet structure. It is the resonating-valence-bond (RVB) state that was proposed by P. W. Anderson \cite{Anderson:87}. We take the non-local pairing correlation to be of the form
\beq
\D_{2}(\tb{e}_l)&\equiv&-\f{2g_2}{N}\sum_{\tb{r}\in\mathcal{A}}\langle a_{\tb{r},\da}b_{\tb{r}+\tb{e}_l,\ua}-a_{\tb{r},\ua}b_{\tb{r}+\tb{e}_l,\da}\rangle,\nn
\eeq
which results in a superconducting order parameter 
\beq
\D_{2,\tb{k}}\equiv \sum_{l=1}^4 \D_2(\tb{e}_l)e^{i\textbf{k}\cdot \tb{e}_l}.\nn
\eeq
Note that this superconducting order parameter is determined by four independent RVB components $\D_{2}(\tb{e}_l)$. 

In the second and third terms of Eq.~(\ref{SC2}), a non-trivial expectation value amounts to a particle-hole correlation $\kappa^\dag=\ex{a_{\tb{r},\ua} b_{\tb{r}+\tb{e}_l,\ua}^\dag +a_{\tb{r},\da} b_{\tb{r}+\tb{e}_l,\da}^\dag}$. The corresponding order parameter is given by 
\beq
\Gamma&\equiv&-\f{g_2}{2N}\sum_{i\in\mathcal{A}}\sum_{l=1}^4\ex{a_{\tb{r},\ua} b_{\tb{r}+\tb{e}_l,\ua}^\dag +a_{\tb{r},\da} b_{\tb{r}+\tb{e}_l,\da}^\dag}.\nn
\eeq 
By performing a mean-field decoupling in the Hamiltonian $H_2$ with respect to the two orders, we obtain
\beq
H_{2,MF}&\simeq&\f{N}{2g_2}\sum_{l=1}^4|\D_2(\tb{e}_l)|^2-\f{2N}{g_2}|\Gamma|^2\nn\\
&&+\sum_\textbf{k} \lt[ \D_{2,\textbf{k}}^\dag (a_{\textbf{k},\da}b_{-\textbf{k},\ua}-a_{\textbf{k},\ua}b_{-\textbf{k},\da})+\tr{H.c.}\rt]\nn\\
&&+\Gamma^\dag \sum_k \gamma_k (a_{k,\ua}^\dag b_{k,\ua}+a_{k,\da}^\dag b_{k,\da})\nn\\&&
-\Gamma \sum_k \gamma_k (a_{k,\ua}b_{k,\ua}^\dag+a_{k,\da}b_{k,\da}^\dag),
\eeq 
where $\gamma_k=\gamma_{-k}=4\cos(k_x d/2)\cos(k_y d/2)$. It is worth noting that the contribution from the second-order particle-hole correlation is negative, as opposed to the contribution from the superconducting order, which can be traced back to an additional overall minus sign when performing the mean-field decoupling for the particle-hole terms in Eq.~(\ref{SC2}).

\subsection{Symmetry of the Order Parameters}\label{symmetry}
Before we discuss the self-consistent procedure to determine the mean-field state, it is important to understand the symmetry of the superconducting order parameter in a basis where the single-particle term is diagonal. By transforming to that basis, the mean-field interaction terms become
\beq
H_{1,MF}&=&\sum_\textbf{k} \D_1^\dag e^{i\theta_{\textbf{k}}}\biggl[ \cos(\theta_{\textbf{k}}) \bigl(\B_{\textbf{k},\da}\B_{-\textbf{k},\ua}+\A_{\textbf{k},\da}\A_{-\textbf{k},\ua} \bigr)\nn\\&&-i\sin(\theta_{\textbf{k}})  \bigl(\B_{\textbf{k},\da}\A_{-\textbf{k},\ua}+\A_{\textbf{k},\da}\B_{-\textbf{k},\ua} \bigr)\biggr]+\tr{H.c.},\nn\\
H_{2,MF}&\simeq&\sum_\textbf{k}  \D_{2,\textbf{k}}^\dag \, e^{i\theta_{\textbf{k}}}\bigl(\B_{\textbf{k},\da}\B_{-\textbf{k},\ua}-\A_{\textbf{k},\da}\A_{-\textbf{k},\ua}\bigr)+\tr{H.c.}\nn
\eeq
Here, $\B_{\textbf{k},\s}$ and $\A_{\textbf{k},\s}$ are the lower and upper band operators, which obey anti-commutation relations, $\theta_{\tb{k}}\equiv\e_{\textbf{k}}/|\e_{\textbf{k}}|$, and the property $\e_{\textbf{k}}=\e_{-\textbf{k}}$ or $\theta_\textbf{k}=\theta_{-\textbf{k}}$ follows from Eq.~(\ref{con}). In $H_{2,MF}$, we have omitted the particle-hole correlation contribution since we are only interested in the symmetry of the superconducting order parameter here. 

In the pairing potential $H_{1,MF}$, we first note that the superconducting order parameter $\D_1$ induces an intra-band pairing $(\B_{-\textbf{k},\ua}^\dag\B_{\textbf{k},\da}^\dag+\A_{-\textbf{k},\ua}^\dag\A_{\textbf{k},\da}^\dag)$ as well as an an inter-band pairing $(\B_{-\textbf{k},\ua}^\dag\A_{\textbf{k},\da}^\dag+\A_{-\textbf{k},\ua}^\dag\B_{\textbf{k},\da}^\dag)$ in the band representation. With the orbital part $ \D_1^\dag e^{i\theta_{\textbf{k}}}$ being an even function in the momentum variable, the intra-band pairing naturally gives rise to a spin singlet structure. For the inter-band pairing, since it is symmetric in the band index, it also results in a spin singlet structure. Thus, the on-site pairing is described by a total order parameter that has a spin singlet structure.

On the other hand, the RVB order $\D_2(\tb{e}_l)$ induces only intra-band pairing $(\B_{\textbf{k},\da}\B_{-\textbf{k},\ua}-\A_{\textbf{k},\da}\A_{-\textbf{k},\ua})$ in $H_{2,MF}$. The spin structure of the pairing is then determined by the parity of the superconducting order $\D_{2,\textbf{k}}$, which depends on the choice of the four-component RVB order. For an even parity $\D_{2,\textbf{k}}=\D_{2,-\textbf{k}}$, it results in a spin-singlet, whereas for an odd parity $\D_{2,\textbf{k}}=-\D_{2,-\textbf{k}}$, it results in a spin triplet. It is important to note that since the external staggered flux breaks the $\mathcal{A}$-$\mathcal{B}$ sublattice symmetry explicitly, the order parameter $\D_{2,\textbf{k}}$ need not have a definite parity. It can generally have a mixed parity with a coherent mixture of states with even and odd parities, and the resulting spin structure is also a coherent mixture of spin singlet and spin triplet states.

\subsection{Mean-Field Free Energy}
The mean-field procedure that we carried out until now involves the introduction of five superconducting order parameters, i.e. $\D_1$ and $\D_2 (\tb{e}_l)$, for $l=1,2,3,4$, and a particle-hole correlation $\Gamma$. For self-consistency, we need to minimize also the resulting free energy. To evaluate the free energy, we first rewrite the grand-canonical mean-field Hamiltonian in the following form
\beq\label{MF2}
H_{MF}=E_0+\sum_{\tb{k}\in 1BZ} \Psi^\dag_{\tb{k}} \mathcal{D}_{\tb{k}}  \Psi_{\tb{k}},
\eeq
with $\Psi^\dag_{\tb{k}}=(a_{\tb{k},\ua}^\dag, b_{\tb{k},\ua}^\dag,b_{-\tb{k},\da},a_{-\tb{k},\da})$,
\beq
E_0=\f{N|\D_1|^2}{g_1}+\f{N}{2g_2}\sum_{l=1}^4|\D_2(\tb{e}_l)|^2-\f{2N}{g_2}|\Gamma|^2,\nn
\eeq
and $\mathcal{D}_{\tb{k}}$ is given by the following matrix
\beq
\label{Matrix}
\bem
-\mu&-\e^*_{\tb{k}}+\Gamma^\dag \gamma_{\tb{k}}&\D_{2,\tb{k}}&\D_1\\
-\e_{\tb{k}}+\Gamma \gamma_{\tb{k}}&-\mu&\D_1&\D_{2,-\tb{k}}\\
\D_{2,\tb{k}}^*&\D_1^*&\mu&\e_{-\tb{k}}^*-\Gamma^\dag \gamma_{-\tb{k}}\\
\D_1^*&\D_{2,-\tb{k}}^*&\e_{-\tb{k}}-\Gamma \gamma_{-\tb{k}}&\mu
\eem.\nn
\eeq
The fermionic chemical potential with equal spin population is $\mu$, and the full lattice dispersion is given by Eq.~(\ref{con}). Since the mean-field Hamiltonian~(\ref{MF2}) is quadratic, we perform a canonical transformation (a Bogoliubov-Valatin transformation) to diagonalize it. The system is then described by two branches of non-interacting fermionic quasi-particles and quasi-holes with spectra $\pm E_{\nu,\tb{k}}$, with $\nu=1,2$. In the new basis, the entropy of the system can be computed simply from that for the ideal Fermi gas
\beq
S&=&-k_B\sum_{\tb{k}} \biggl\{f(E_{\nu,\textbf{k}})\ln f(E_{\nu,\textbf{k}})\nn\\&&+[1-f(E_{\nu,\textbf{k}})]\ln [1-f(E_{\nu,\textbf{k}})]\biggr\},\nn
\eeq
where $f(E_{\nu,\textbf{k}})=1/[\exp(\B E_{\nu,\textbf{k}})+1]$ is the Fermi-Dirac distribution. The free energy is then given by
\beq\label{free}
F\bigl[\D_1,\D_2(\tb{e}_l)\bigr]&=&E-TS\nn\\
&=&E_0-\f{1}{\B}\sum_{\nu=1}^2\sum_{\textbf{k}}\biggl\{\ln \lt(1+e^{-\B E _{\nu,\textbf{k}}}\rt)\nn\\&&+\ln \lt(1+e^{\B E_{\nu,\textbf{k}}}\rt)\biggr\}.
\eeq
We see that the $|\Gamma|^2$-term renders the free energy unbounded below, which is an artifact of the mean-field decoupling for a particle-hole correlation. Indeed, the mean-field procedure is not suited to treat such a correlation. We will therefore ignore its contribution and set it to zero $\Gamma=0$ for the rest of the work. In the following sections, we minimize the above expression with respect to the variational parameters $(\D_1,\D_2(\tb{e}_l))$ and identify the regime of parameters where superconductivity can occur.

\begin{figure}[t]
\subfigure[]{
\includegraphics[scale=.65, angle=0, origin=c]{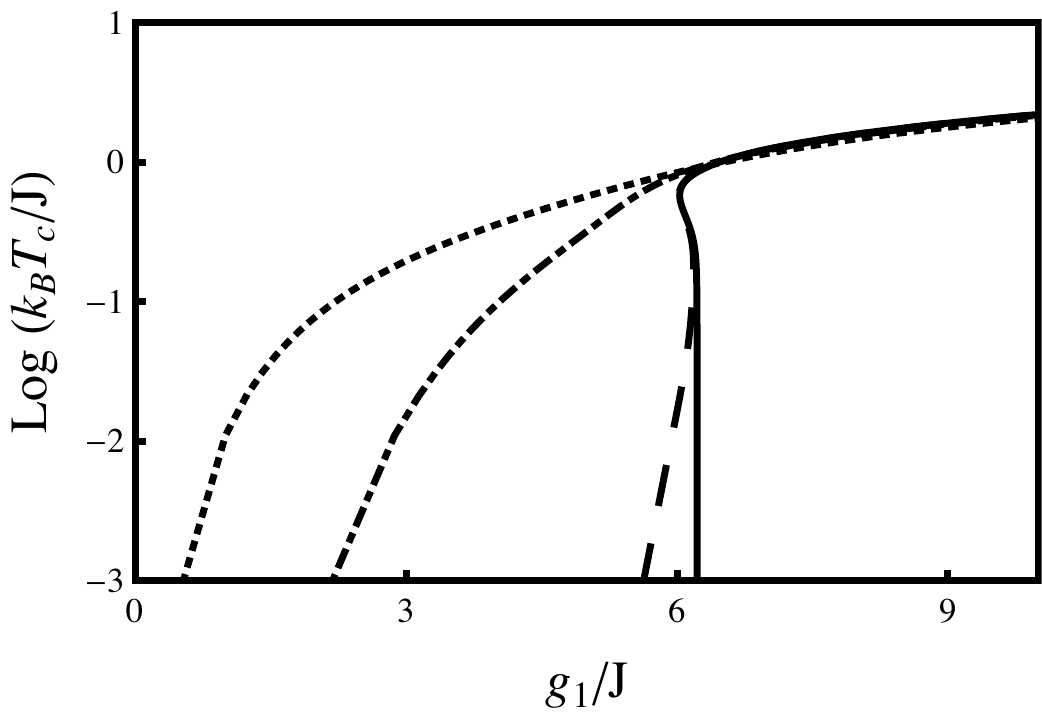}}\hspace{1cm}
\subfigure[]{
\includegraphics[scale=.65, angle=0, origin=c]{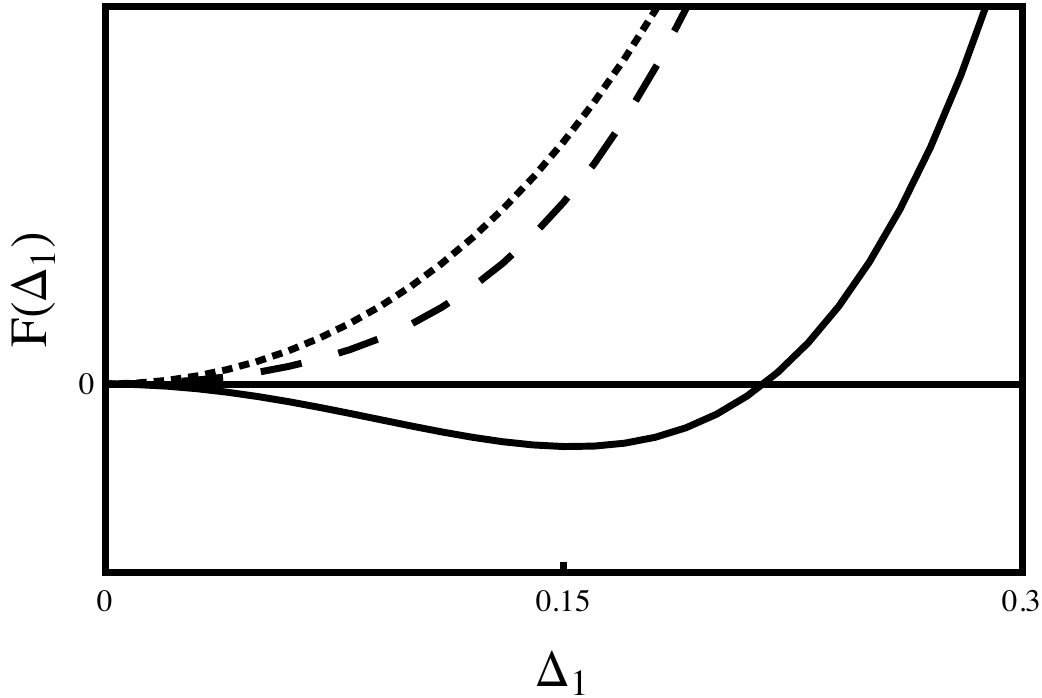}}\caption{\label{fig.2} (a) Solutions to the gap equation for local superconductivity at the phase transition for different chemical potentials: $\mu/J= 0$ (solid), $\mu/J=0.1$ (dash), $\mu/J=1$ (dotted dash), $\mu/J=2$ (dotted). (b) To show the reentrant behavior, we plot the evolution of the free energy at different temperatures $k_B T/J=0.8 \tr{\ (dashed),\ } 0.56 \tr{\ (solid),\ } 0.1 \tr{\ (dotted),\ }$ for a fixed coupling $g_1/J=6.1$ and chemical potential $\mu/J=0$. }
\end{figure}

\subsection{Local Superconductivity}
In this section we consider only the local pairing channel characterized by the order parameter $\D_1$. Setting $\D_{2,\tb{k}}=0$, the quasi-particle spectra can be readily obtained,
\beq
E_{1,2,\textbf{k}}&=&\sqrt{|\e_\textbf{k}|^2+\mu^2+\D^2_1\mp2 \sqrt{|\e_\textbf{k}|^2\mu^2+(\tr{Im}\,\e_\textbf{k})^2 \D^2_1}}.\nn
\eeq
By substituting these into the free energy function Eq.~(\ref{free}) and extremizing the latter with respect to $\D_1$,
\beq
\f{\partial F(\D_1)}{\partial \D_1}=0,
\eeq
we obtain the gap equation
\beq
1&=&\f{g_1}{4N}\sum_{\nu=1}^2\sum_\textbf{k}  \biggl\{\f{\tanh\lt(\B E_{\nu,\textbf{k}}/2\rt)}{E_{\nu,\textbf{k}}}\nn\\&&\times\biggl[ 1+\f{(-1)^{\nu}(\tr{Im}\,\e_\textbf{k})^2}{\sqrt{|\e_\textbf{k}|^2 \mu^2+(\tr{Im}\,\e_\textbf{k})^2\D_1^2}}\biggr]\biggr\}.
\eeq
To determine the second-order phase transition between the normal and the superconducting phases, we take the limit $\D_1\rightarrow 0$ in the gap equation where the superconducting gap vanishes smoothly. The resulting equation
\beq
1&=&\f{g_1}{4N}\sum_\textbf{k} \biggl\{ \f{\tanh
\bigl[\B_c(|\e_\textbf{k}|-\mu)/2\bigr]}{|\e_\textbf{k}|-\mu}\lt[ 1-\f{(\tr{Im}\,\e_\textbf{k})^2}{|\e_\textbf{k}| |\mu|}\rt]\nn\\&&+ \f{\tanh\bigl[\B_c(|\e_\textbf{k}|+\mu)/2\bigr]}{|\e_\textbf{k}|+\mu}\lt[ 1+\f{(\tr{Im}\,\e_\textbf{k})^2}{|\e_\textbf{k}| |\mu|}\rt]\biggr\}
\eeq
determines the critical temperature $\B_c=1/k_B T_c$ for a given coupling $g_1$ and chemical potential $\mu$, see Fig.~\ref{fig.2}(a).

We first note that for zero chemical potential, the Fermi level lies at the conical points of the Dirac cone. Due to the vanishing of the density of states, the system is quantum critical. This means that the system can undergo a phase transition even at zero temperature, where there are no thermal fluctuations. In the present case, it is a second-order phase transition driven purely by quantum fluctuations. The quantum critical point, which is found to be $g_{1,c}/J\simeq 6.2$ in this channel, separates the normal and the superconducting phases. On the other hand, for finite chemical potential the system ceases to be quantum critical. The usual BCS picture of Fermi surface instability is then recovered where an infinitesimal attractive interaction favors superconductivity.
%%can one check the T_c dependence on the coupling? is it an exponential?

Secondly, for small chemical potential $\mu/J\ll 1$, we observe that the $T_c$-curve displays a non-monotonous dependence on the coupling around the temperature region $k_B T_c/J\sim \mathcal{O}(1)$. This is due to the intrinsic nature of the inter-band pairing in this channel. Inter-band pairing generally requires more energy fluctuations, since there is an energy gap in the pairing between a state from the upper energy band and its time-reverse partner from the lower energy band. For vanishingly small chemical potential, we are left with two energy scales in the problem, namely the thermal energy and the coupling. Whenever thermal fluctuations are suppressed for $k_B T/J\lesssim 1$, a stronger coupling is therefore required to promote the pairing. Thus, we see the increase in the coupling strength required to induce superconductivity as thermal fluctuations are cut off around the region $k_B T/J\sim \mathcal{O}(1)$. This feature is called a re-entrant behavior because even when the system is in the symmetry-broken (ordered) phase below the critical temperature, it can get back to the disordered phase by lowering further the temperature. This is only valid for a coupling around the critical value $g_{1,c}$ and a small chemical potential. To confirm the re-entrant behavior, we also investigate the free energy function at various temperatures for a fixed coupling and zero chemical potential, as shown in Fig.~\ref{fig.2}(b). The absolute minimum of the free energy indeed returns to the origin where the system is disordered, as the temperature is lowered.

\subsection{Extended Superconductivity}
Now, we want to study the extended superconducting channel. In this case, we set $\D_1=0$ in the Hamiltonian (\ref{MF2}). Furthermore, for the sake of simplicity, we assume the four-component RVB order to be real. We then arrive at the following properties for the superconducting order parameter:
\beq
\D_{2,\textbf{k}}&=&\D_{2,-\textbf{k}}^*,\nn\\
|\D_{2,\textbf{k}}|^2&=&\sum_l\D_2(\tb{e}_l)^2\nn\\&&+\sum_{l\neq m}\D_2(\tb{e}_l)\D_2(\tb{e}_m) \cos(\textbf{k}\cdot(\tb{e}_l-\tb{e}_m)),\nn\\
\tr{Im}\,\D_{2,\textbf{k}}&=&\sum_l \D_2(\tb{e}_l) \sin(\textbf{k}\cdot \tb{e}_l).\nn
\eeq
The quasi-particle spectra are then given by
\beq
E_{1,2,\textbf{k}}&=&\sqrt{|\e_\textbf{k}|^2+|\D_{2,\textbf{k}}|^2+\mu^2\mp2 |\e_\textbf{k}|\sqrt{(\tr{Im}\,\D_{2,\textbf{k}})^2+\mu^2}}.\nn
\eeq
Upon extremizing the free energy function
\beq
\f{\partial}{\partial \D_2(\tb{e}_l)} F[\D_2(\tb{e}_1),\D_2(\tb{e}_2),\D_2(\tb{e}_3),\D_2(\tb{e}_4)]=0,
\eeq
for $l=1,2,3,4$, we obtain a set of four coupled gap equations
\beq
\D_2(\tb{e}_l)&=&\f{g_2}{2N}\sum_{\textbf{k}}\biggl\{\lt(\f{\tanh\lt(\B E_{1,\textbf{k}}/2\rt)}{E_{1,\textbf{k}}}+\f{\tanh\lt(\B E_{2,\textbf{k}}/2\rt)}{E_{2,\textbf{k}}}\rt)
 \nn\\&&\times\biggl(\D_2(\tb{e}_l)+\sum_{m\neq l}\D_2(\tb{e}_m)\cos[\textbf{k}\cdot(\tb{e}_l-\tb{e}_m)]\biggr)\nn\\&&+\biggl(-\f{\tanh\lt(\B E_{1,\textbf{k}}/2\rt)}{E_{1,\textbf{k}}}+\f{\tanh\lt(\B E_{2,\textbf{k}}/2\rt)}{E_{2,\textbf{k}}}\biggr)\nn\\&&\times\f{|\e_\textbf{k}|\sin(\textbf{k}\cdot \tb{e}_l)\,\tr{Im}\,\D_{2,\textbf{k}}}{\sqrt{(\tr{Im}\,\D_{2,\textbf{k}})^2+\mu^2}}\biggr\}
\nn
\eeq
for $l=1,2,3,4$. We now take the system to be close to the phase transition, where the gap $\D_2(\tb{e}_l)$ is small, and expand the right-hand side of the gap equations to leading order in the gap to obtain the linearized gap equations\begin{figure}
\subfigure[]{
\includegraphics[scale=.65, angle=0, origin=c]{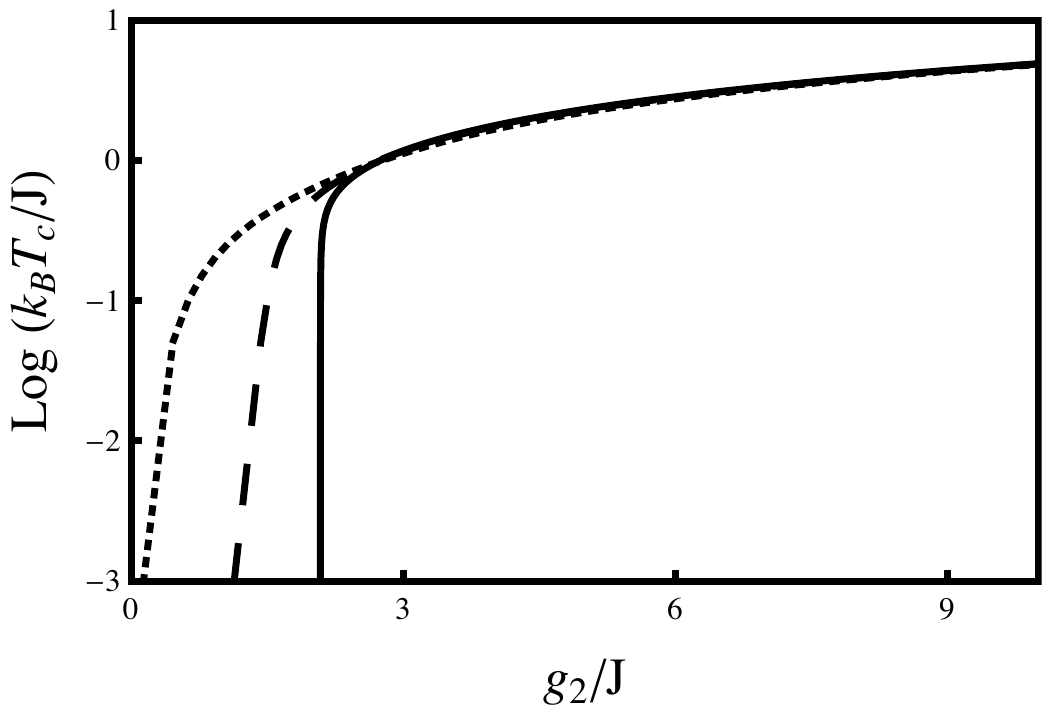}}
\hspace{0.5cm}
\subfigure[]{
\includegraphics[scale=0.8, angle=0, origin=c]{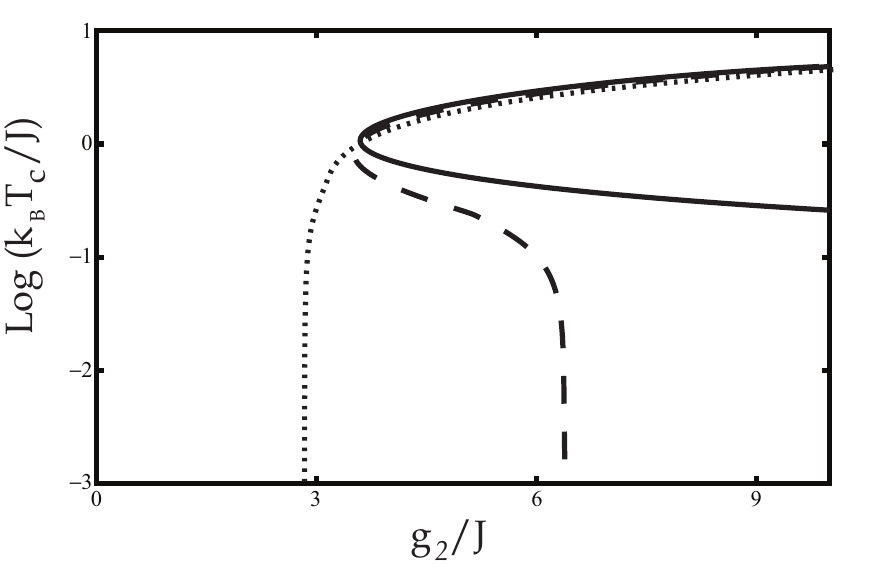}}\caption{\label{fig.3}Solutions to the gap equation with (a) $d$-wave symmetry and (b) $p$-wave symmetry at the phase transition for different chemical potentials: $\mu/J\rightarrow 0$ (solid), $\mu/J=1$ (dashed), and $\mu/J=2$ (dotted).}
\end{figure}
\beq\label{linearized}
\D_2(\tb{e}_l)&=&\f{g_1}{2N}\sum_{m=1}^4\sum_{\textbf{k}} \biggl[
\biggl(\f{\tanh[\B_c( |\e_\textbf{k}|+|\mu|)/2]}{|\e_\textbf{k}|+|\mu|}\nn\\&&\!\!\!\!\!\!
+\f{\tanh[\B_c( |\e_\textbf{k}|-|\mu|)/2]}{|\e_\textbf{k}|-|\mu|} \biggr)
\cos[\textbf{k}\cdot \tb{e}_l]\cos[\textbf{k}\cdot \tb{e}_m ]\nn\\&&\!\!\!\!\!\!+\f{\sinh(\B_c |\mu|)}{|\mu| \cosh[\B_c( |\e_\textbf{k}|+|\mu|)/2]\cosh[\B_c( |\e_\textbf{k}|-|\mu|)/2]}\nn\\&&\!\!\!\!\!\!\times\sin (\textbf{k}\cdot \tb{e}_l) \sin (\textbf{k}\cdot \tb{e}_m)\biggr]\D_2(\tb{e}_m)
\eeq
for $l=1,2,3,4$. The linearized gap equations can also be written in the matrix form
\beq
\f{1}{g_2}\Om=\bem
D&B&C&B\\B&D&B&C\\C&B&D&B\\B&C&B&D
\eem\Om,\nn
\eeq
where the four-vector $\Om$ is defined by $\Om^T\equiv(\D_2(\tb{e}_1),\D_2(\tb{e}_2),\D_2(\tb{e}_3),\D_2(\tb{e}_4))$, and $D,B,C$ are matrix elements obtained from the coupled Eqs.~(\ref{linearized}). In fact, a numerical evaluation shows that the matrix element $B\simeq 0$. It is clear now that the problem of determining the critical temperature $T_c$ amounts to finding non-trivial solutions to the matrix equation. The eigenvalues and their corresponding eigenvectors are classified as follows:
\begin{align}\label{dwave}
\frac{1}{g_2} =  \left\{
\begin{array}{ll}
C+D & :(1,0,1,0),(0,1,0,1), \ \ d{\rm-wave}\\
-C+D & :(-1,0,1,0),(0,-1,0,1), \ \ p{\rm-wave}
\end{array}
\right.
\end{align}
with
\beq
C+D&=&\f{1}{2N}\sum_\textbf{k} 2\biggl\{ \f{\tanh[\B_c( |\e_\textbf{k}|+|\mu|)/2]}{|\e_\textbf{k}|+|\mu|}\nn\\&&
+\f{\tanh[\B_c( |\e_\textbf{k}|-|\mu|)/2]}{|\e_\textbf{k}|-|\mu|}\biggl\} \cos^2(\textbf{k}^+ d),\nn\\
-C+D&=&\f{1}{2N}\sum_\textbf{k} 2 \f{\sinh(\B_c |\mu|)}{|\mu| \cosh[\B_c( |\e_\textbf{k}|+|\mu|)/2]}\nn\\&&\times\f{\sin^2(\textbf{k}^+ d)}{\cosh[\B_c( |\e_\textbf{k}|-|\mu|)/2]},\nn
\eeq
where $\textbf{k}^{\pm} =(k_x\pm k_y)/2$. The four eigenvectors come in two classes of symmetries, which reflect the underlying irreducible representation of the symmetry group of the square lattice. In Fig.~\ref{fig.3}, we solve for the critical temperature as a function of the coupling, for the two classes of eigenvectors. Again, a quantum critical behavior is expected for zero chemical potential. However, in contrast to the local pairing in the previous section, this channel does not involve inter-band pairing and the system should not display re-entrant features. In Fig.~\ref{fig.3}(a) we show the $T_c$-curves for the $d$-wave channel which behave as expected. The quantum critical coupling is $g_{2,c}/J\simeq 2.1$, which is smaller than the critical local pairing coupling.

\begin{figure}
\subfigure[]{\label{fig.4a}
\includegraphics[scale=.6, angle=0, origin=c]{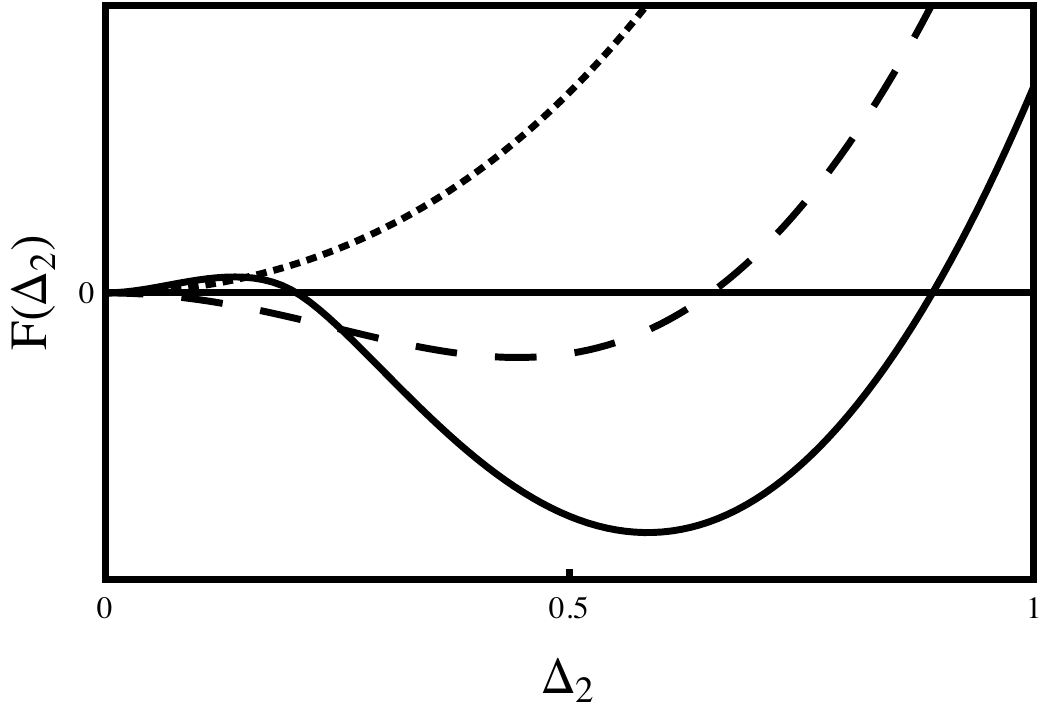}}
\hspace{.8cm}
\subfigure[]{\label{fig.4b}
\includegraphics[scale=.6, angle=0, origin=c]{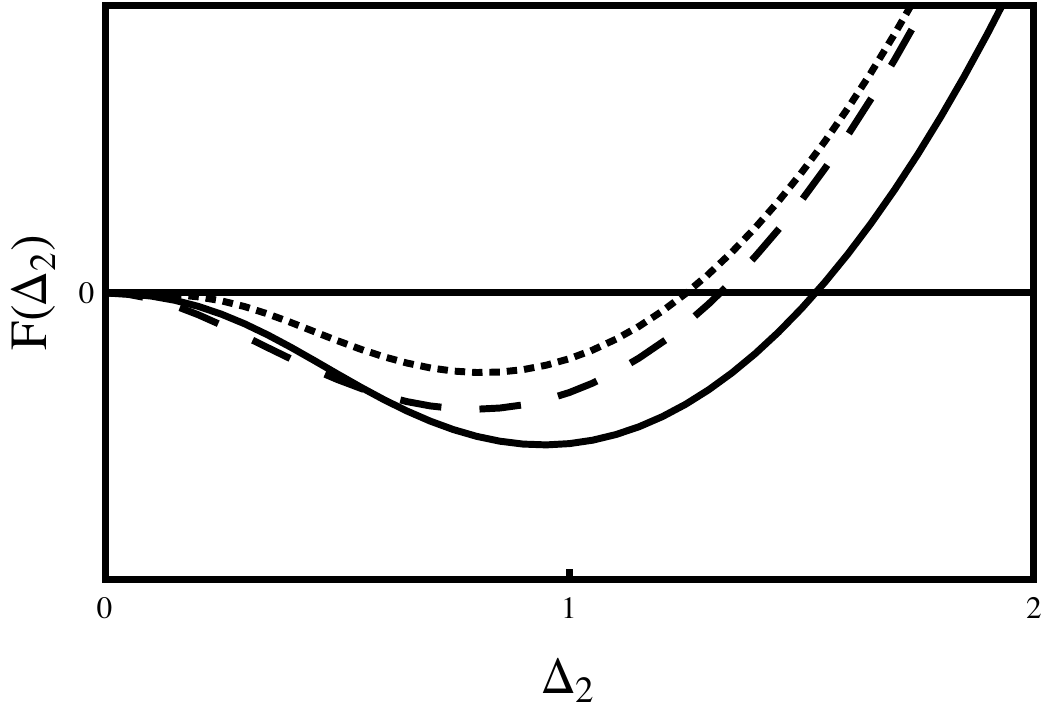}}
\vspace{.3in}
\caption{\label{fig.4}(a) Evolution of the free energy for the $p$-wave superconductivity at different temperatures $k_B T/J=3 \tr{\ (dotted),\ } 1.5 \tr{\ (dashed),\ } 0.1 \tr{\ (solid),\ }$ at a fixed coupling $g_2/J=5$ and chemical potential $\mu/J=0$. (b) Comparing the free energy for the $p$-wave channel $(-\D_2/\sqrt{2},0,\D_2/\sqrt{2},0)$ (dotted), the $d$-wave channel $(\D_2/\sqrt{2},0,\D_2/\sqrt{2},0)$ (dashed), and the mixed channel $(\D_2,0,0,0)$ (solid) which yields the lowest free energy, for $\mu=0$, $k_B T/J=0.5$, $g_2/J=5$.}
\end{figure}
On the other hand, the $T_c$-curves for the $p$-wave channel exhibit highly irregular features, see Fig~\ref{fig.3}(b). To understand that this is an artifact of the solution of the gap equation, let us analyze the free energy function in the $p$-wave channel more closely. By decreasing the temperature below the critical value given by the upper part of the $T_c$-curve at a fixed coupling, the change in the free energy is shown in Fig~\ref{fig.4}(a). While the absolute minimum is shifted away from the origin, signaling that the system enters the symmetry-broken phase, a new local minimum is also being developed at the origin below the second critical temperature. Since the linearized gap equation is an expansion around the origin, the lower part of the $T_c$-curve seen in Fig~\ref{fig.4}(b) merely describes the development of the new local minimum. We thus conclude that there is no re-entrant behavior. However, the development of another local minimum at the origin gives rise to the possibility of a first-order phase transition where the $T_c$-curve develops a vertical slope. Looking at the free energy, we indeed find a first-order phase transition line starting at a tricritical point, where it meets the upper part of the second-order phase transition $T_c$-curve.

Given the choices of pairing states with different symmetries, we need to determine the most favorable one by comparing the free energy deep in the superconducting phase. Comparing the $T_c$-curves for the two channels in Fig.~\ref{fig.3}(a) and Fig.~\ref{fig.3}(b), we see that the $d$-wave channel generally has a higher critical temperature for coupling $g_2/J\lesssim 2.8$. For $g_2/J \gtrsim 2.8$, since the $p$-wave channel can be favorable, we need to consider more general states with mixed symmetry. As already discussed in Sect.~\ref{symmetry}, the lack of parity symmetry for the pairing state in this channel allows for a coherent mixing of order parameters with different symmetries. By taking the four eigenvectors as a basis which spans the space of the order parameter, we look for the vector which yields the lowest free energy. As shown in Fig.~\ref{fig.4}(b), we find the state $(1,0,0,0)$ to be the most favorable (lowest free energy) in the superconducting phase and the corresponding order parameter reads
\beq
\D_{2,\textbf{k}}=\D_2 \biggl[ \cos\biggl(\f{k_x+k_y}{2}\,d\biggr)+i\sin \biggl(\f{k_x+k_y}{2}\,d\biggr) \biggr].
\eeq
A comparison of the real-space configurations of the local and non-local pairing, see Figs.~\ref{Fig.5}(a) and \ref{Fig.5}(b), shows that the latter leads to unconventional superconductivity, where both the gauge  and the $C_{4v}$ crystal lattice symmetries are spontaneously broken.

\begin{figure}
\subfigure[]
{\label{Fig.5}\includegraphics[scale=.25, angle=0, origin=c]{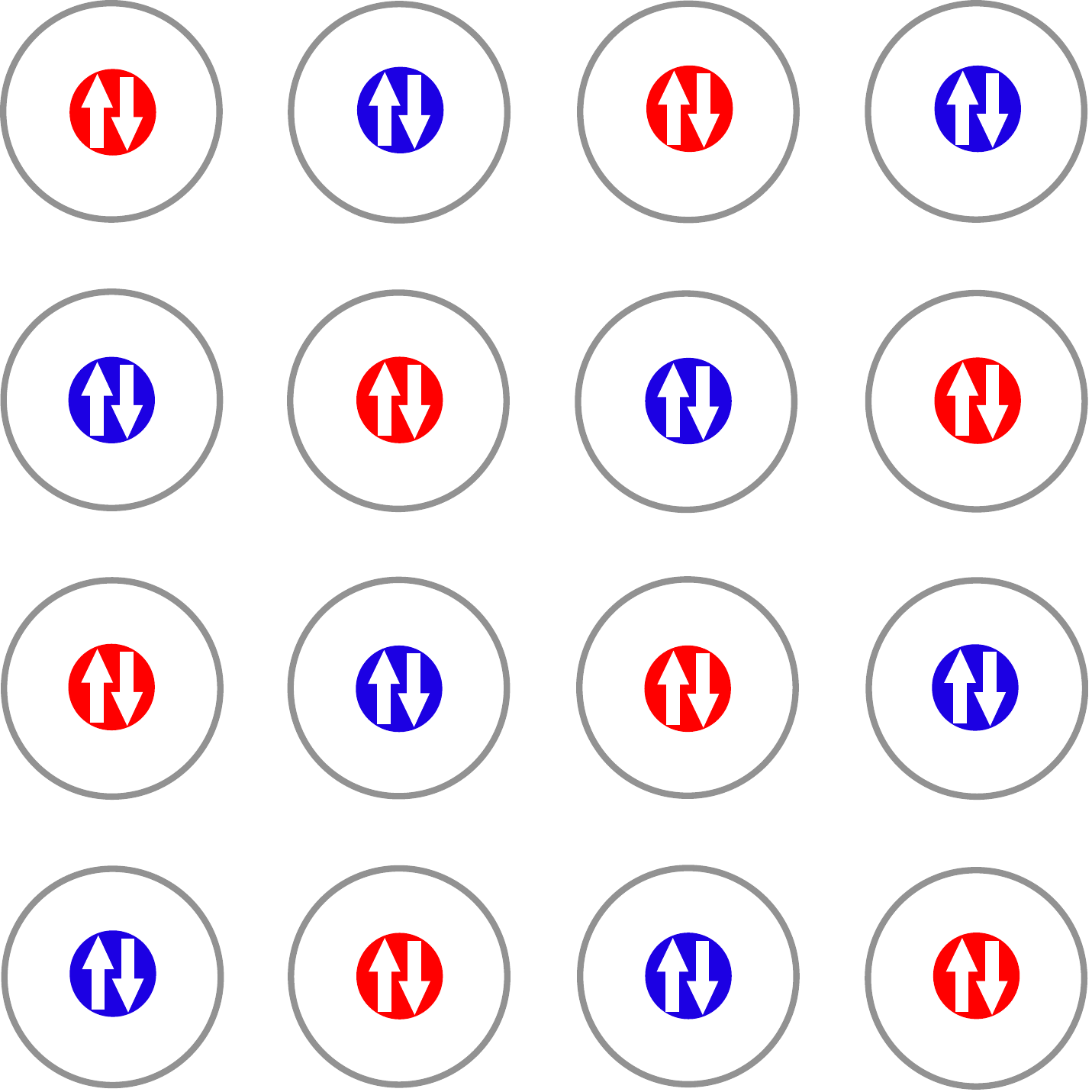}}
\subfigure[]{\includegraphics[scale=.25, angle=0, origin=c]{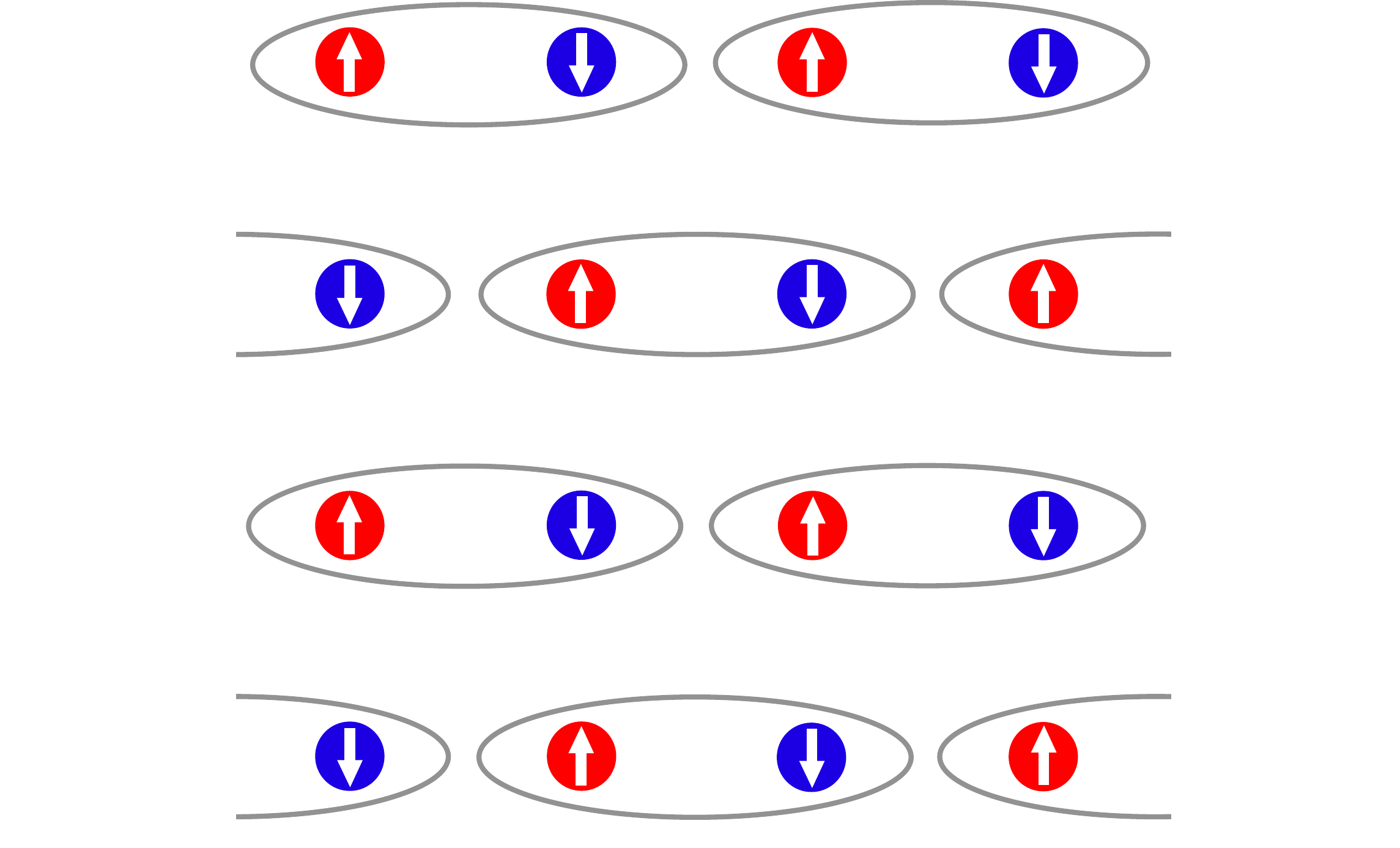}}
\caption{ \label{singlet}(color online) (a) Real-space configuration of the local pairing $s$-wave superconducting phase.  (b) Real-space configuration of the nearest-neighbor spin-singlet bonding with one non-vanishing component.}
\end{figure}

\subsection{Phase Diagram at Quantum Criticality}
Until now we treated the two superconducting channels separately. Here we consider the competition between the different types of superconductivity at the quantum critical regime with zero chemical potential and a staggered flux $\p=\pi$. The discussion for general values of chemical potentials and flux values will be presented in the next section.

\begin{figure}
\includegraphics[scale=1, angle=0, origin=c]{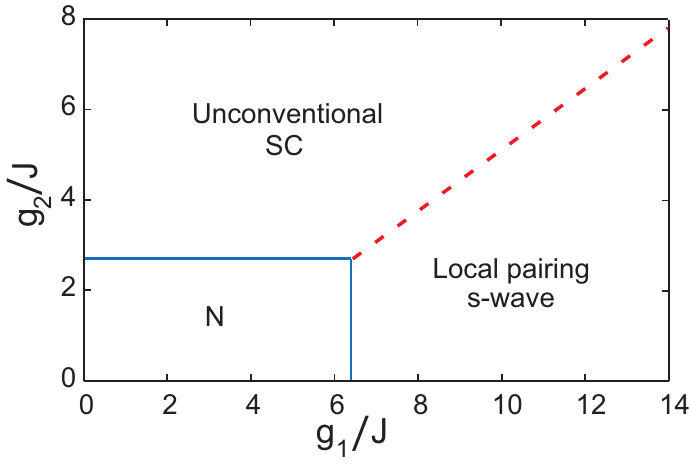}
\caption{ \label{free_en}(color online) Zero-temperature mean-field phase diagram for the Hamiltonian (\ref{interactingHAmiltonian}) with zero chemical potential and a staggered flux $\p=\pi$. The two superconducting phases are separated by a first order phase transition line (dash line) between them, and a second order phase transition (full line) with the normal phase. }
\end{figure}

We have shown that the most favorable superconducting state in the separate channels are the constant gap $\D_1$ and the unconventional superconducting state $\D_2(\tb{e}_l)=(1,0,0,0)$, respectively. We choose the latter, even though we have already noted that the extended superconducting channel has a much richer behavior around the region $g_2/J \simeq 2.8$. By taking into account the possibility of co-existence between the two superconducting phases, we numerically minimize the free energy $F(\D_1,\D_{2,\textbf{k}})$ containing both superconducting orders. By locating the global minimum in the free energy, we find an unpaired phase (normal phase), an $s$-wave superconducting phase and a non-local pairing superconducting phase in the phase diagram, see Fig.~\ref{free_en}. The two superconducting phases are separated by a first-order phase transition (dashed line) with no region of coexistence. A multi-critical point is identified at $(g_1/J,g_2/J)=(6.2,2.1)$ at zero temperature, where the first-order phase transition line and the two second-order lines meet.

It is important now to justify the validity of the mean-field approach to the superconductivity problem. The general criterion for the validity of the mean-field theory in the weak coupling regime is that the coupling strength must be smaller than the energy bandwidth $\delta E$. For the system we study, the energy bandwidth is given by
\beq
\delta E=4J\sqrt{2+2\cos\biggl(\f{\phi}{2}\biggr)}.
\eeq
The weak coupling regime is then given $g_{1,2}< \d E$. Referring to the phase diagram in Fig.~\ref{free_en} and a bandwidth $\d E\simeq5.7 J$, we find that both the $g_1$ and $g_2$ channels are approximately within the weak coupling regime, even though the $g_1$ channel is closer to an intermediate coupling regime. Nonetheless, for coupling strength within the weak coupling regime, the three distinct many-body phases are already accessible.

\subsection{Phase Diagram Away from Quantum Criticality}
In this section, we consider arbitrary values of the chemical potential. In this case, rather than the chemical potential, the physical quantity that can be controlled in experiments is the particle density, or the fermion filling fraction $\langle n \rangle$, the average number of spin-$1/2$ fermions per lattice site. Furthermore, comparing the two sets of solution for the gap equations in the different channels at zero chemical potential, see Fig.~\ref{fig.2}(a) and Fig.~\ref{fig.3}(b), we note that the unconventional superconducting channel yields the highest transition temperature $T_c$. Thus, we shall consider only this channel with a varying fermion filling fraction. At the phase transition $T=T_c$, where the superconducting gap vanishes, the fermion filling fraction $\langle n \rangle$ can be determined quite easily by the non-interacting limit (ignoring the Hartree energy)
\beq\label{doping}
\d&=&|\langle n \rangle-1|=\f{1}{2N}\sum_{\tb{k}}\biggl[ \tanh\lt(\f{|\e_{\tb{k}}|+|\mu|}{2 k_B T_c}\rt)\nn\\&&-\tanh\lt(\f{|\e_{\tb{k}}|-|\mu|}{2 k_B T_c}\rt)\biggr],
\eeq
where $N$ is the total number of sites. The quantity $\d$ is conventionally called hole (particle) doping in solid state materials, since it measures the departure of the electronic density from the half-filling (particle-hole symmetric) limit. By solving the doping equation~(\ref{doping}) self-consistently with the linearized gap equation~(\ref{linearized}), we obtain the $T_c$-curve as a function of doping summarized in Fig.~\ref{Fig.5}.

\begin{figure}\includegraphics[scale=.55, angle=0, origin=c]{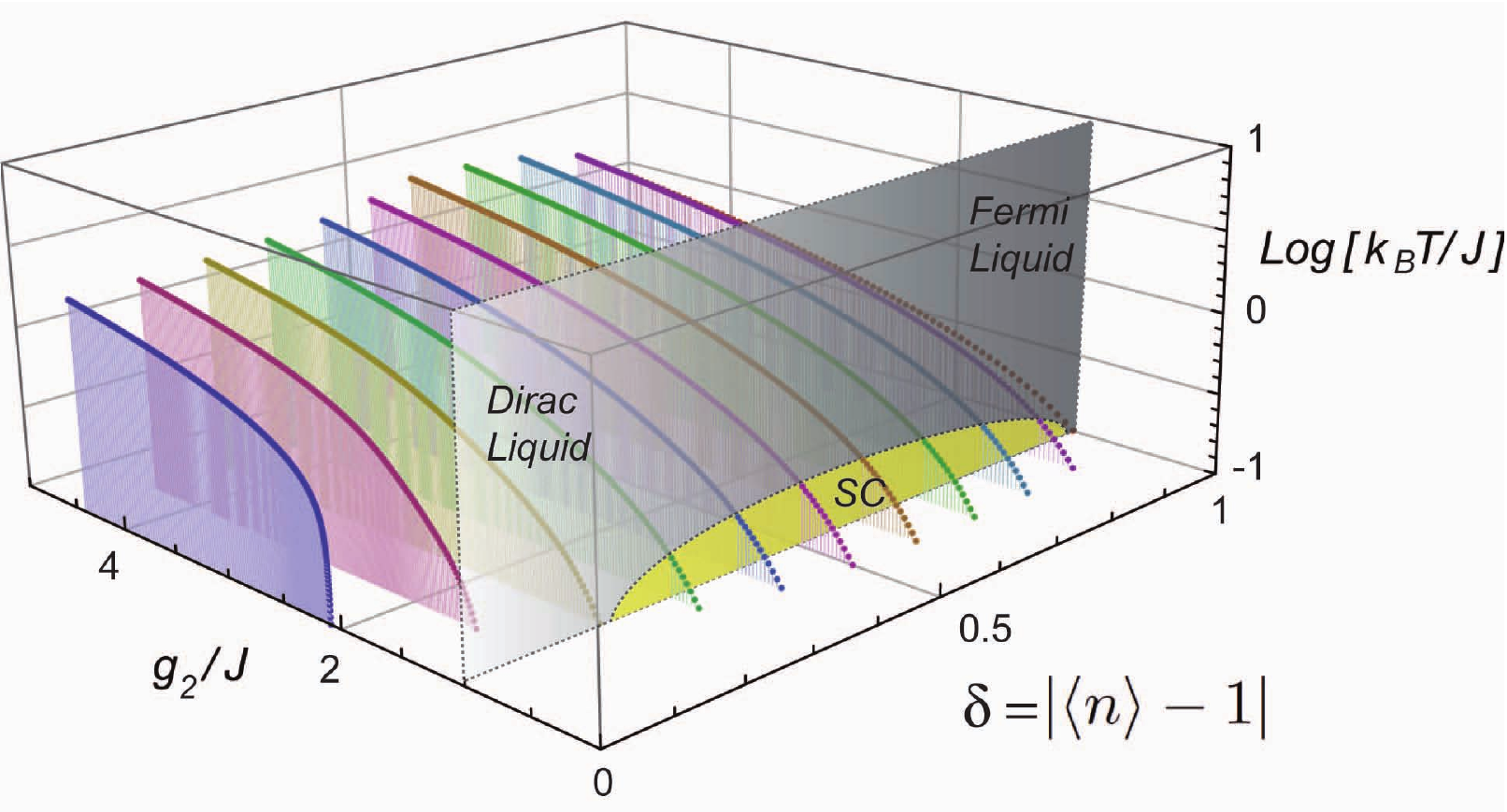}\caption{\label{Fig.5}(color online) Finite-temperature phase diagram in the unconventional superconducting channel for different doping $\delta$, for $g_1/J<6.2$. The shaded plane at a fixed coupling $g_2/J=1$ shows a dome-like shape superconducting phase embedded in the normal phase. The latter interpolates Dirac liquid behavior at low-doping to a Fermi liquid behavior at high-doping.}\end{figure}

We see that as soon as the system is tuned away from unit filling fraction ($\mu=0$, $\delta=0$), the system ceases to be quantum critical (The exponential tail in the $T_c$-curves extending to down to $g_2/J=0$ is however not visible in the temperature scale of Fig~\ref{Fig.5}).  Secondly, the shift of the $T_c$-curves is not a monotonous function of the filling fraction. The $T_c$ value increases as the filling fraction decreases from unity, but below a filling fraction of approximately 0.3, the $T_c$ value decreases again (see the $T_c$-curves for different filling fractions in Fig.~\ref{Fig.5}).      

%While the dispersion remains linear (Dirac-like) as the filling fraction $\langle n \rangle$ is slightly tuned away from unity ($\mu=0$, $\delta=0$), the system ceases to be quantum critical. \red{The $T_c$ versus $g_2$ graph develops an exponential tail extending all the way to $g_2=0$ (not visible for the scale used in Fig.~\ref{Fig.5}) and shifts towards lower values of the coupling strength $g_2$. This shift, however, does not grow monotonously as the filling fraction is further decreased but reduces again, below a filling fraction of approximately 0.3 ($\delta \approx 0.7$ in Fig.~\ref{Fig.5}).}

\begin{figure*}
\includegraphics[scale=1]{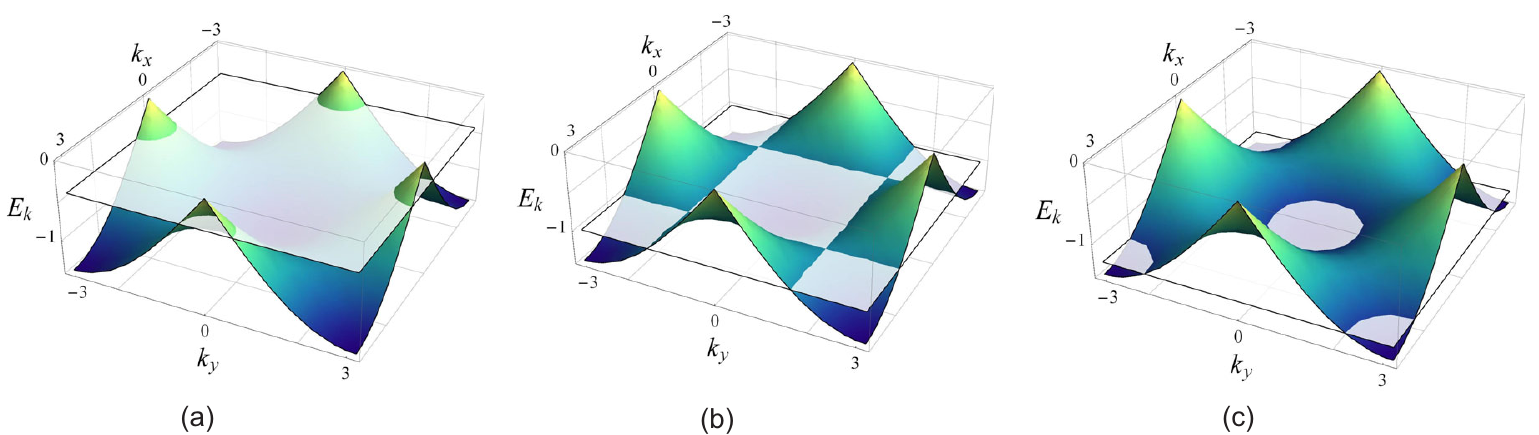}
\caption{\label{Fig.5b}(color online) The intersection of the shaded plane with the lower branch single-particle spectrum represents the topology of the Fermi pockets. The filling fraction determines the height of the shaded plane. From figures (a)-(c) the filling fraction is decreasing.}
\end{figure*}

To understand better the non-monotonous shift of the $T_c$-curves, it is illustrative to observe the evolution of the non-interacting Fermi surface. As shown in Fig.~\ref{Fig.5b}, as we tune the doping from zero to unity, the nature of the fermionic carriers changes from hole-like to particle-like at the doping $\d=0.5$, or the quarter-filling fraction. For $\d<0.5$, the Fermi wave vector $k_F$ increases in the two inequivalent Fermi pockets as the doping increases. For $\d>0.5$ the Fermi wave vector $k_F$ decreases instead as the doping continues to increase. Qualitatively speaking, the non-monotonous shift of the $T_c$-curve is based on the behavior of the Fermi wave vector. The BCS picture relates the transition temperature to the Fermi wave vector via an exponential function, $T_c \sim e^{-1/k_F |a|}$, where $a$ is proportional to the scattering length in an atomic system. This means that for an increasing Fermi wave vector in the doping region $\d\in [0,1/2]$, there will be an increasingly higher transition temperature, while for the decreasing Fermi wave vector in the doping region $\d\in [0,1/2]$, the transition temperature decreases.

Finally, by plotting temperature versus doping for a fixed value of the coupling strength $g_1$, as shown in the shaded plane in Fig.~\ref{Fig.5}, we find a dome-shaped unconventional superconducting phase at intermediate filling fractions, surrounded by the normal phase for fillings close to zero or unity, which we termed Dirac-liquid (Fermi-liquid) on the left (right) side of the phase diagram, where linear (quadratic) single-particle dispersion prevails. The dome-structure of the unconventional superconducting phase is similar to the phase diagram for high-$T_c$ cuprates and heavy fermions.

\subsection{Away from Isotropic Dirac Cones}
\begin{figure}
\includegraphics[scale=0.8, angle=0,  origin=c]{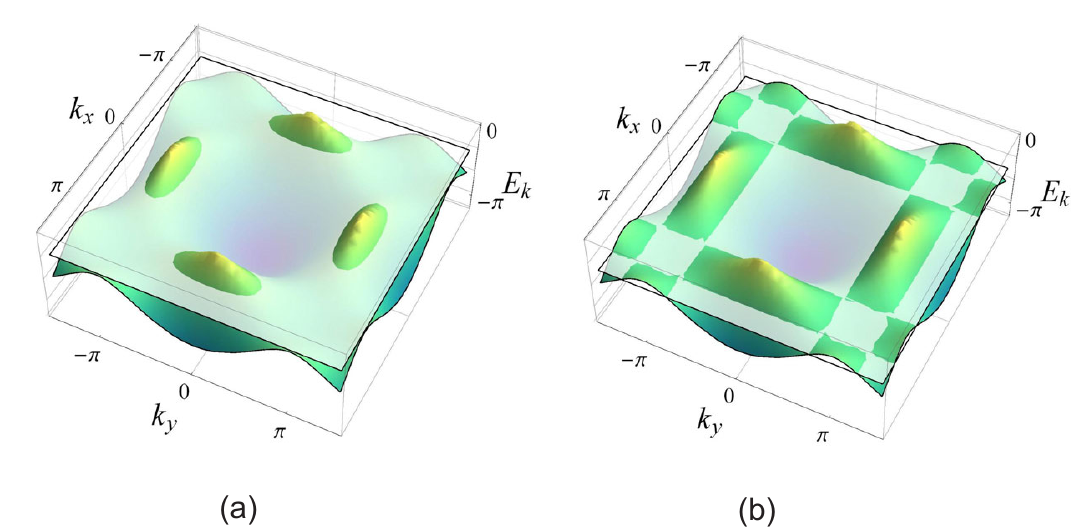}
\caption{ \label{Fig.6}(color online) (a) For flux $\phi\neq\pi$, the Fermi pockets generally take the shape of ``banana" due to the anisotropic Dirac cones. (b) At lower filling fraction, the resulting sets of nesting vector are also different from the $\pi$-flux case due to Fermi ``squares" of different sizes. }\end{figure}

So far we have fixed the staggered flux value to be $\phi=\pi$. Except for the special values of $\phi= 2\pi\nu$, $\nu\in\mathbb{Z}$ we do not expect a qualitative change in the mean-field results. The effect of the different staggered flux values is a change in the anisotropy of the Dirac cones. For a finite doping, it results in anisotropic Fermi surfaces, see Fig.~\ref{Fig.6}.

\section{Experimental Considerations}
The superconductivity considered in this article arises for temperatures on the  order of ten percent of the Fermi-temperature. This temperature range  can be accessed in state of the art Bose-Fermi mixtures subjected to  an optical lattice. A possible experimental scenario is to employ the  widely used rubidium-potassium system composed of a balanced mixture of fermionic $^{40}$K-atoms prepared in the $|F=9/2,m_F=-7/2\rangle$  and $|F=9/2,m_F=-9/2\rangle$ Zeeman components of the $F=9/2$ ground  state hyperfine level and bosonic $^{87}$Rb atoms in the $|F=1,m_F=1 \rangle$ ground state \cite{RbK}. The parameter $U_{BF}^2/U_ {BB}$ may be adjusted via its dependence on the well depth, while an  $s$-wave Feshbach resonance around 202 Gauss can be used to tune $U_ {FF}$ independently. In order to experimentally discriminate the two  SC phases discussed here, one could search for a signature of their  distinct gap functions in their momentum spectra \cite{Ket:06}.  Correlation measurements similar to the one described in Ref.~\cite {Jin:05} should be another powerful method to obtain information on  the nature of the pairing. 

\section{Discussions and Conclusions}
We considered here an ultracold Bose-Fermi mixture in a 2D square optical lattice subjected to an effective staggered magnetic field, which exhibits a Dirac-like single-particle spectrum. The system is studied at low-temperatures, such that the bosons condense and mediate a longer-range (nearest neighbor) attractive interaction between the fermions. At half-filling, the Dirac fermions experience both, local (Hubbard) and nearest neighbor attractive interactions, which can be tuned independently. The zero-temperature mean field phase diagram exhibits a competition between a local $s$-wave and a non-local unconventional superconducting phases. It is interesting to compare the superconductivity occurring here to that of graphene-like systems \cite{Zhao:06,Ghaemi:07,Bergman:09}. In the square lattice with a staggered flux, equivalent Dirac cones are related to each other by time-reversal. In contrast, in the graphene lattice, the time-reversal operation maps one Dirac cone to the other inequivalent Dirac cone. Thus, when BCS Cooper pairs are formed in the square geometry, no different flavors of Dirac fermions are involved. This is not the case for the hexagonal symmetry.

At finite temperatures, the appearance of unconventional superconductivity below a dome reveals an intriguing link to strongly correlated electronic materials. In addition, the evolution of the normal phase (surrounding the superconducting dome) from a Dirac-liquid to a Fermi-liquid upon increasing doping is another essential feature of high-$T_c$ cuprates and heavy fermions, not easily captured by usual theoretical descriptions of the system. Finally, the appearance of ``banana-shaped" Fermi-pockets for flux values $\phi\neq\pi$, reminiscent of the ones observed in high-$T_c$ cuprates by means of ARPES experiments~\cite{bananas}, adds to the number of puzzling similarities. The local anisotropies in the Fermi surface in Fig.~\ref{Fig.6} could eventually lead to striped ground states, like those observed in the cuprates~\cite{TranquadaKivelson}. The multitude of similarities between the system considered here and the high-$T_c$ superconductors suggests that also for high-$T_c$-materials a generalized Hubbard model with complex rather than real hopping coefficients might be the appropriate description. This speculation presumes that some physical mechanism could be breaking the time-reversal symmetry already in the pseudogap phase of cuprates, which is in fact supported by recent observations \cite{Kapitulnik}. A remaining open question is then to identify which particle should be playing the role of the bosons in the cuprates, to mediate a longer-range interaction, which ultimately leads to unconventional superconductivity. In any case, cold atomic systems are already proving to be a fascinating playground to develop our understanding of high-Tc superconductors. 

%A long list of candidates could be envisioned here. However direct the correspondence may be, cold atomic systems prove as a very useful test ground for increasing our understanding of high-$T_c$-superconductivity.

\section*{Acknowledgements} This work was partially supported by the Netherlands Organization for Scientific Research (NWO).
AH acknowledges support by DFG (He2334/10-1) and Joachim Herz Stiftung.

\section{Appendix}\label{app1}
Here we provide the details of the Bogoliubov approximation procedure for bosonic operators to arrive at the mean-field action Eq.~(\ref{fluc}). The procedure amounts to selecting out the condensation mode and making the replacement $\B_{\tb{k}_0}\rightarrow\sqrt{N_0}+\B_{0}$ in the Hamiltonian. The Bogoliubov approximation amounts to keeping only the terms with fluctuation operators up to second order. For convenience, we define the various coupling constants as $a_1\equiv U_{BB}/4N,\, a_2\equiv U_{BF}/N$. For the terms containing only bosonic operators we have\begin{widetext}
\beq
H_B&=&\sum_\textbf{k} (E_\textbf{k}-\mu)\B_\textbf{k}^\dag\B_\textbf{k}+a_1\!\!\sum_{\textbf{k}_1,\textbf{k}_2,\textbf{k}_3,\textbf{k}_4}\!M(\textbf{k}_1,\ldots,\textbf{k}_4) \B^\dag_{\textbf{k}_1}\B_{\textbf{k}_2}^\dag\B_{\textbf{k}_3}\B_{\textbf{k}_4}\d(\textbf{k}_1+\textbf{k}_2-\textbf{k}_3-\textbf{k}_4)\nn\\
&\simeq&(E_0-\mu)N_0+\biggl[(E_0-\mu)+4a_1N_0\,\biggr]\sqrt{N_0}(\B_0+\B_0^\dag)
+2 a_1  N_0^2+\sum_\textbf{k} (E_\textbf{k}-\mu+8a_1N_0)\B_\textbf{k}^\dag\B_\textbf{k}\nn\\&&+a_1 N_0\sum_\textbf{k} M(\textbf{k}_0,\textbf{k}_0,\textbf{k},-\textbf{k})\B_\textbf{k}\B_{-\textbf{k}}+a_1 N_0\sum_\textbf{k} M(\textbf{k},-\textbf{k},\textbf{k}_0,\textbf{k}_0)\B_\textbf{k}^\dag\B_{-\textbf{k}}^\dag,\nn
\eeq
where we have used $M(\textbf{k}_1,\textbf{k}_2,\textbf{k}_1,\textbf{k}_2)=2$. And for the boson-fermion interaction term on the $\mathcal{A}$-sublattice we get
\beq
H_{BF,\,\mathcal{A}}&=&a_2\sum_{\tb{r}\in \mathcal{A}} \sum_\s n_{\tb{r},\s}n_{\tb{r}}^B=a_2\sum_{\tb{r}\in \mathcal{A}} \sum_\s\lt\{n_{\tb{r},\s}\sum_{\textbf{k}_1} g^*_{\textbf{k}_1} \B^\dag_{\textbf{k}_1}e^{-i\textbf{k}_1\cdot \tb{r}}\sum_{\textbf{k}_2}g_{\textbf{k}_2}\B_{\textbf{k}_2}e^{i\textbf{k}_2\cdot \tb{r}}\rt\} \nn\\
&=&a_2\sum_{\tb{r}\in \mathcal{A}} \sum_\s \biggl\{\biggl(\tilde{n}_{\s} + n_{\tb{r},\s}\biggr)\biggl[ N_0 +\ro{N_0} (\B_0^\dag+\B_0) +\ro{N_0} \,g^*_0\, e^{-i\textbf{k}_0\cdot \tb{r}}\sum_\textbf{k} g_\textbf{k}\B_\textbf{k} e^{i\textbf{k}\cdot \tb{r}} \nn\\&&+\ro{N_0} g_0 e^{i\textbf{k}_0\cdot \tb{r}}\sum_\textbf{k} g^*_\textbf{k}\B_\textbf{k}^\dag e^{-i\textbf{k}\cdot \tb{r}}+ \sum_{\textbf{k}_1,\textbf{k}_2}g_{\textbf{k}_1}^*g_{\textbf{k}_2}\B_{\textbf{k}_1}^\dag\B_{\textbf{k}_2}e^{-i\textbf{k}_1\cdot  \tb{r}+i\textbf{k}_2\cdot \tb{r}}\biggr]\biggr\}\nn\\
&\simeq&a_2  N_0 N\sum_\s \tilde{n}_{\s} +a_2 \ro{N_0}N \sum_\s \tilde{n}_{\s}  (\B_0+\B^\dag_0)+a_2N\tilde{n}_{\s} \sum_\textbf{k} \B_\textbf{k}^\dag \B_\textbf{k}+a_2 N_0 \sum_{\tb{r}\in \mathcal{A}}\sum_{\s} n_{\tb{r},\s}\nn\\&&+a_2\ro{N_0}\sum_{\tb{r}\in \mathcal{A}}\sum_{\s} n_{\tb{r},\s} g_0^*e^{-i\textbf{k}_0\cdot \tb{r}}\sum_\textbf{k} g_\textbf{k} \B_\textbf{k} e^{i\textbf{k} \cdot \tb{r}}+a_2\ro{N_0}\sum_{\tb{r}\in \mathcal{A}}\sum_{\s} n_{\tb{r},\s} g_0 e^{i\textbf{k}_0\cdot \tb{r}}\sum_\textbf{k} g_\textbf{k}^*\B_\textbf{k}^*e^{-i\textbf{k} \cdot \tb{r}},\nn
\eeq
where we have introduced a mean-field density $\tilde{n}_{\s}$ for the spin $\s$ fermions and thus, fluctuation terms of higher order, i.e., $\mathcal{O} \,(n_{\tb{r},\s} \B^\dag_\textbf{k}\B_\textbf{k})$, can be neglected. Performing the same procedure for the boson-fermion interaction term on the $\mathcal{B}$-sublattice yields a similar expression $H_{BF,\mathcal{B}}$. Collecting all the terms we get
\beq
&&H_B+H_{BF,\,\mathcal{A}}+H_{BF,\mathcal{B}}\nn\\&&=\biggl(E_0-\mu+ U_{BB} n_0/2+U_{BF}\sum_\s \tilde{n}_{\s} \biggr)N_0+\sqrt{N_0}\biggl(E_0-\mu+U_{BB}n_0+U_{BF} \sum_\s \tilde{n}_{\s}\biggr)(\B_0+\B_0^\dag)\nn\\&&
+\sum_\textbf{k} \biggl(E_\textbf{k}-\mu+2 U_BB n_0+U_{BF} \sum_\s \tilde{n}_\s\biggr)\B_\textbf{k}^\dag\B_\textbf{k}+\f{1 }{4}U_{BB} n_0\sum_\textbf{k} M(\textbf{k}_0,\textbf{k}_0,\textbf{k},-\textbf{k})\B_\textbf{k}\B_{-\textbf{k}}\nn\\&&+\f{1}{4}U_{BB} n_0\sum_\textbf{k} M(\textbf{k},-\textbf{k},\textbf{k}_0,\textbf{k}_0)\B_\textbf{k}^\dag\B_{-\textbf{k}}^\dag+U_{BF}\,n_0 \sum_{\tb{r}\in\mathcal{A}}\sum_{\s} (n_{\tb{r},\s}+n_{\tb{r}+\tb{e}_1,\s})\nn\\
&&+U_{BF}\ro{\f{n_0}{2N}}\sum_\textbf{k} \biggl\{\sum_{\tb{r}\in\mathcal{A}} \sum_{\s} n_{\tb{r},\s}  g_0^* g_\textbf{k} e^{-i(\textbf{k}_0-\textbf{k})\cdot\tb{r}}+\sum_{\tb{r}\in\mathcal{A}}\sum_{\s} n_{\tb{r}+\tb{e}_1,\s}  e^{-i(\textbf{k}_0-\textbf{k})\cdot(\tb{r}+\tb{e}_1)}\biggr\} \B_\textbf{k}\nn\\
&&+U_{BF}\ro{\f{n_0}{2N}}\sum_\textbf{k} \biggl\{\sum_{\tb{r}\in\mathcal{A}} \sum_{\s} n_{\tb{r},\s}  g_0 g_\textbf{k}^* e^{i(\textbf{k}_0-\textbf{k})\cdot\tb{r}}+\sum_{\tb{r}\in\mathcal{A}} \sum_{\s} n_{\tb{r}+\tb{e}_1,\s}  e^{i(\textbf{k}_0-\textbf{k})\cdot(\tb{r}+\tb{e}_1)}\biggr\} \B_\textbf{k}^\dag.
\eeq
\end{widetext}

\end{document}